\documentclass{cas-sc}
\usepackage[backend=bibtex,sorting=none,maxcitenames=1,maxbibnames=99,style=numeric-comp]{biblatex}
\usepackage[utf8]{inputenc}
\usepackage[tiny]{titlesec}
\usepackage{url}
\usepackage{hyperref}\hypersetup{pdfborder={0 0 0}}
\usepackage{blindtext}
\usepackage{amsmath}
\usepackage{array}
\usepackage{tabularx,booktabs}
\usepackage{multirow}
\usepackage{amssymb}
\usepackage{graphicx}
\usepackage{wrapfig}
\usepackage{float}
\usepackage{caption,subcaption}
\usepackage[colorinlistoftodos]{todonotes}
\usepackage{tcolorbox}
\usepackage{bm}
\usepackage{cleveref}
\usepackage{algorithm,algcompatible}
\usepackage{algpseudocode}
\usepackage{cancel}
\usepackage{miller}

\DeclareMathOperator{\sgn}{sgn}

\definecolor{C0}{HTML}{1F77B4}
\definecolor{C1}{HTML}{FF7F0E}
\definecolor{C2}{HTML}{2ca02c}
\definecolor{C3}{HTML}{d62728}
\definecolor{C4}{HTML}{9467bd}
\definecolor{C5}{HTML}{8c564b}

\makeatletter
\newlength{\bibsep}{\@listi \global\bibsep\itemsep \global\advance\bibsep by\parsep} 
\makeatother
\ifdefined\usetodonotes
\usepackage[colorinlistoftodos]{todonotes}
\fi

\ifdefined\usechanges
\usepackage{fontawesome}
\usepackage[commentmarkup=todo,authormarkup=none]{changes} 
\definechangesauthor[color=C0]{R1} 
\definechangesauthor[color=C1]{R2} 
\makeatletter
\renewcommand{\Changes@Markup@comment}[3]{%
  \IfStrEq{\Changes@optioncommentmarkup}{todo}%
		{\colorlet{Changes@todocolor}{authorcolor}\todo[color=Changes@todocolor!10, bordercolor=Changes@todocolor, linecolor=Changes@todocolor!70, nolist]{\faBookmarkO\   \textsf{\textbf{#1}}}}{}}
\makeatother
\newtcolorbox{addedbox}[3][]
{
  colframe=#2,
  colback=#2!10,
  title={\faBookmarkO\ \textsf{\textbf{#3}}},
  left=0pt,right=0pt,
  #1,
}
\else
\usepackage{xparse}
\NewDocumentCommand{\added}{ o o m }{#3}
\NewDocumentCommand{\deleted}{ o o m }{}
\NewDocumentCommand{\replaced}{ o o m m }{#3}
\NewDocumentEnvironment{addedbox}{ m m }{}{}
\fi

\def\F{\bm{\mathrm{F}}}
\def\I{\bm{\mathrm{I}}}
\def\P{\bm{\mathrm{P}}}
\def\Fgb{\mathrm{\Delta}\bm{\mathrm{F}}^{\mathrm{GB}}}

\def\areavec{\bm{\mathrm{a}}}
\def\nor{\bm{\mathrm{n}}}
\def\A{\mathrm{A}}
\def\G{\mathrm{G}}
\def\verts{\mathbb{V}(\graph)}
\def\edges{\mathbb{E}(\graph)}
\def\area{a}
\def\vol{V}
\def\etal{{\it et al.}}

\begin{document}

\shorttitle{}
\shortauthors{}
\title[mode=title]{Grain boundary network plasticity: reduced-order modeling of deformation-driven shear-coupled microstructure evolution}
\author[uccs]{Daniel Bugas}
\author[isu,uccs]{Brandon Runnels}[orcid=0000-0003-3043-5227]
\cormark[1]
\cortext[1]{Corresponding author}
\ead{brunnels@iastate.edu}
\address[uccs]{Department of Mechanical and Aerospace Engineering, University of Colorado, Colorado Springs, CO USA}
\address[isu]{Department of Aerospace Engineering, Iowa State University, Ames, IA USA}


\begin{abstract}
    Microstructural evolution in structural materials is known to occur in response to mechanical loading and can often accommodate substantial plastic deformation through the coupled motion of grain boundaries (GBs).
    This can produce desirable behavior, such as increased ductility, or undesirable behavior such as mechanically-induced coarsening.
    In this work a novel, multiscale model is developed for capturing the combined effect of plasticity mediated by multiple GBs simultaneously.
    This model is referred to as ``grain boundary network plasticity.''
    The mathematical framework of graph theory is used to describe the microstructure connectedness, and the evolution of microstructure is represented as volume flow along the graph.
    By using the principle of minimum dissipation potential, which has previously been applied to grain boundary migration, a set of evolution equations are developed that transfer volume and eigendeformation along the graph edges in a physically consistent way.
    It is shown that higher-order geometric effects, such as the pinning effect of triple points, may be accounted for through the incorporation of a geometric hardening that causes geometry-induced GB stagnation.
    The result is a computationally efficient reduced order model that can be used to simulate the initial motion of grain boundaries in a polycrystal with parameters informed by atomistic simulations.
    The effectiveness of the model is demonstrated through comparison to multiple bicrystal atomistic simulations, as well as a select number of GB engineered and non-GB engineered data obtained from the literature.
    The effect of the network of shear-coupling grain boundaries is demonstrated through mechanical response tests and by examining the yield surfaces.
\end{abstract}

\maketitle

\section{Introduction}

Microstructural evolution can produce great variability in material properties, and can result from thermal \cite{gokuli2021multiphase,o2016exploration}, mechanical \cite{oddershede2011grain,tourret2022phase}, electric \cite{chakraborty2018phase}, chemical \cite{elsener2009variable,vikrant2020electrochemical}, and radiative loadings \cite{campana2008grain, borovikov2013coupled}.
Predicting how grain boundaries (GB) evolve is essential to mitigating material failure by improving understanding of how microstructural evolution impacts material properties' evolution.
It has been shown experimentally \cite{weiss1998grain, molodov2015grain,rupert2009experimental} and through atomistic simulations \cite{thomas2017reconciling,thomas2019disconnection} that microstructure responds mechanistically through GB motion to accommodate applied stress with shear coupling \cite{cahn2006coupling, watanabe1999control, chen2019grain}.
As a GB moves under stress, the swept volume shears with the GB motion called shear coupling.
The ensemble of shear coupled GB migration in a microstructural network effects a plastic response, accumulating permanent deformation mediated by the motion of the boundaries.

\added[id=R1]{
  The plastic response of crystalline materials is historically understood to be accomplished through the motion of dislocations.
  This is particularly true in large-grained polycrystalline materials and in crystalline structures with numerous available slip systems.
  Indeed, for many decades, the primary role of grain boundaries was understood to be as an inhibitor (or, in some cases, a source) of dislocations, resulting in the well-known Hall-Petch effect.
  However, with increasing attention turned towards nanocrystalline and thin-film materials, it has become apparent that GBs act as plasticity mediators themselves.
  This was shown in the seminal work of \textcite{shan2004grain}, and confirmed through numerous experimental studies showing the role of GB plasticity (and grain growth) in nanocrystalline aluminum \cite{jin2004direct} and nanocrystalline copper \cite{zhang2004influence}, and even some microstructures with larger grain sizes \cite{mompiou2009grain}.
  Thin films were also observed to be structures in which GB mediated plasticity is a dominant deformation mechanism, as reported in \cite{gianola2006stress,lohmiller2013effect,mompiou2013inter}.
  Remarkably, TEM analysis showed that nanoscale thin films were absent of dislocations, despite exhibiting properties traditionally understood to be mediated by dislocations \cite{haque2004deformation}.
  This has led to growing consensus that ductility in thin films is mediated largely through grain boundary mechanisms, i.e. sliding and shear coupling \cite{liebig2021grain}.
  Similar observations have been made regarding deformation processes in nanocrystalline metals: \textcite{lohmiller2014untangling} concluded that dislocation plasticity contributes only about 40\% to overall deformation in nanocrystalline nickel.
}

\added[id=R1,comment={1.1}]{ 
  The mounting evidence of the importance of GB-mediated mechanisms in thin-film and nanocrystalline materials indicates that they must be accounted for in computational modeling efforts.
  While great strides have been made in the understanding of GB deformation mechanisms,
  (the reader is referred to \textcite{han2018grain} and \textcite{rollett2021grain} for comprehensive reviews),
  computational work at larger scales has been primarily concentrated on dislocation-based plasticity.
  This motivates the need for a method with a direct treatment of GB plasticity, as informed by current knowledge of GB deformation mechanisms, with sufficient reduction in order so as to be tractable at continuum length scales.
  We propose such a model in this work.
  For simplicity as well as clarity of presentation, we consider GB plasticity mechanisms {\it only}, noting that for systems where GB mechanisms dominate (e.g. \cite{lohmiller2014untangling}), this is a reasonable simplification.
  Therefore, the presentation here leaves dislocation effects to be accounted for as a natural, but future, extension.
}

GBs are generally understood to respond in one of four ways under an applied stress \cite{cahn2004unified}: (1) GB migration normal to the GB plane, (2) GB migration normal to the GB plane with an additional shear in the swept region by the GB, (3) grain sliding parallel to GB plane, (4) and grain rotation in which the lattice misorientation changes with respect to the GB. 
Mechanism 2 is characterized by the shear coupling constant, $\beta$, which is dictated by the crystallography of the GB, and which coupling represents the ratio of parallel and tangential motion, $\beta=\frac{v_\parallel}{v_\perp}$.
Mechanisms 1-3 can then be unified by understanding mechanism 1 and mechanism 3 to be the limiting cases of mechanism 2, corresponding to $\beta\to0$ and $\beta\to\infty$ (sliding), respectively.
All three can then be modeled in the context of mechanism 2, where sliding is approximated with a large but finite value for $\beta$.
Therefore, except for the effect of grain rotation, all relevant grain boundary mechanisms can be modeled as instances of shear coupling.

At an atomistic level, GB migration is generally understood to be mediated though the creation, propagation, and annihilation of ``disconnections:'' step defects in a boundary that includes a step height ($h$) and a Burger's vector ($\mathbf{b}$) \cite{rajabzadeh2014role,kvashin2020atomic}. 
The shear coupling factor is related to disconnection type through the relationship $\beta=|\mathbf{b}|/h$, and both are determined by the bicrystallography and character of the boundary.
It is possible to enumerate the (countably infinite) disconnection modes systematically using the coincident site lattice (CSL) and displacement shift complete lattice (DSCL) for symmetric tilt grain boundaries (STGB) and, more recently, for asymmetric tilt grain boundaries (ATGB) as well \cite{admal2022interface}.
However, the coupling factor itself is not sufficient to determine the motion of the boundary \cite{chen2019grain}: temperature and loading conditions can generate substantially different behaviors in the same boundary.
Many low $\Sigma$ boundaries at moderate temperatures have GB motion driven by minimizing the distance traveled by individual atoms \cite{chesser2021optimal}, but  non-optimal distance minimizing reshuffling may not occur at high temperatures, producing the ejection of defects \cite{chesser2022taxonomy}.
Current knowledge of grain boundary shear coupling is primarily driven by atomistic simulations \cite{yu2021survey,yu2019survey}, much of which has been confirmed by experiment \cite{mompiou2009grain,mompiou2011direct}.
However, atomistic modeling techniques are inherently limited in length and time scales, making it impossible to determine the long-time behavior of large-scale microstructures.
This necessitates a scale-bridging approach.

A number of mesoscale approaches have been advanced towards modeling the disconnection-mediated motion of shear coupled boundaries.
A continuum model for GB migration driven by underlying disconnections was originally proposed in \cite{zhang2017equation}, and successfully applied to polycrystals \cite{zhang2021equation}, triple-junction motion \cite{thomas2019disconnection,wei2020grain}, and multi-mode migration using the phase field method \cite{han2022disconnection,salvalaglio2022disconnection}.
This approach provides a much-needed update to traditional curvature-driven boundary migration.
Admal \etal\ \cite{admal2018unified,he2021polycrystal}, demonstrated that the Kobayashi-Warren-Carter (KWC) model for crystal grain evolution can be used to model shear coupled boundary motion using geometrically necessary dislocations (GND), with the benefit that GBs integrate naturally with dislocations, and can interact with them without special treatment.
A similar formulation was also proposed by \cite{mikula2019phase}, also based on the KWC framework.
An alternative phase field model, proposed in \cite{runnels2020phase,gokuli2021multiphase}, showed that disconnections arise as natural mediators of grain boundary migration, when grain boundary anisotropy is considered along with nonlinear kinetics (thresholding). 
These mesoscale methods offer valuable insight into the behavior of boundaries at larger length and time scales than is possible through atomistics, but are still inherently limited in scope due to the degree of resolution required.
To efficiently simulate a large-scale structure, additional reduction of order is necessary.

A standard workhorse in the computational mechanics of polycrystalline materials is crystal plasticity finite elements (CPFE) \cite{abdolvand2015study,aoyagi2014crystal,wong2016crystal,choi2009microstructure,ma2006consideration,roters2010overview}.
CPFE effects coarse-grained simulation of dislocation behavior through the reduction of order from individual dislocations to average accumulated plastic slip.
By replacing dislocation dynamics with coarse-grained flow laws, it is possible to account for the unique mechanical behavior of polycrystals without prohibitive computational cost.
Strain gradient crystal plasticity \cite{wulfinghoff2013gradient,han2005mechanism,van2013grain,bayley2006comparison,voyiadjis2019strain} makes it possible to account for nonlocal effects and GNDs, and introduces a fundamental length scale.
Crystal plasticity has been used to study the interaction of dislocations with GBs \cite{evers2004scale,ma2006consideration,wulfinghoff2013gradient,dahlberg2013deformation}, recovering the Hall-Petch effect \cite{evers2004scale}.
A limitation of most CPFE approaches, however, is the difficulty associated with modeling GB migration.
The sharp interface model by Basak and Gupta \cite{basak2016plasticity} considers bulk plasticity between coupled discontinuous defects, and a set of governing equations for sharp interfaces can be used to evolve a CPFE mesh \cite{lazar2011more,eren2022comparison}.
However, sharp interface motion requires the consideration of topological transitions \cite{eren2021topological}, making efficient microstructure simulation at appreciable scales difficult.
There remains a need for a modeling framework that is similar to crystal plasticity, but able to account for permanent deformation mediated by the evolution of microstructure.

To capture the effect of GB shear coupling on large-scale mechanics, it is necessary to model shear coupling using the theoretical framework of plasticity.
This was proposed by Berbenni \etal\ \cite{berbenni2013micromechanics}, who developed a micromechanics-based model for bicrystals that accurately replicates boundary stick slip behavior.
More recently, Chesser \etal\ proposed a thermodynamic framework for general modeling of GB migration in a continuum sense, using the principle of minimum dissipation potential \cite{chesser2020continuum}.
By comparison to atomistic simulations, it was shown that a dissipation potential effectively encapsulates the intrinsic GB properties corresponding to migration behavior, and that the minimization of dissipation potential can serve as a mechanism for mode selection.
An advantage of this approach, as with most variational models, is its invariance with respect to choice of internal variables: whereas the original formulation was in terms of discrete interface position, \cite{gokuli2021multiphase} demonstrated that an order parameter can be used to generate a phase field model.

In this work, a continuum-level model is proposed to describe the aggregate mesoscale effects of GB motion in a highly reduced-order framework.
  The idea of homogenizing the behavior of a polycrystalline aggregate is not new.
  Grain-to-continuum averaging had been proposed nearly a century ago by Taylor \cite{taylor1938plastic} and Sachs \cite{sachs1928ableitung}, whose models are still in used successfully today to estimate plastic response of randomly-oriented polycrystals.
  In this work, the goal is to capture the plastic evolution of an entire network of microstructure, mediated through the shear-coupled motion of grain boundaries (``disconnection-mediated plasticity'') an a manner similar to that of the classical Taylor and Sachs models.
  Unlike these approaches, however, a model capturing grain boundary shear-coupling
must capture the nuances of texture and grain connectedness via reduced order parameters, without resolving microstructure texture explicitly. 
This requires a structure to facilitate basic communication between neighboring grains, and encode physical parameters of each grain, such as volume, shear, stress, etc.
The purpose of this work is to develop a model that satisfies these constraints.
It should be emphasized that the framework allows for the inclusion of additional mechanisms (such as crystal plasticity) that are not considered here. 
Therefore, the framework as presented in this work is most suitable to materials in which GB motion is the dominant mechanism for deformation, such as nanocrystalline materials or polycrystals with a hexagonal close-pack (HCP) structure.

The remainder of the paper is structured in the following way.
\Cref{sec:theory} presents the graph representation of microstructure, and associated reduced-order kinematics, that are necessary to describe microstructure evolution by GB migration.
Governing equations (balance of momentum) and kinetic laws (principle of minimum dissipation potential) are presented in the NP context, and exact solutions are derived.
Higher order microstructural effects, such as triple point pinning and GB stagnation, are accounted for using hardening potentials.
\Cref{sec:examples} presents a number of examples demonstrating the behavior of the NP model.
It should be emphasized that the choices of examples were driven primarily by the availability of data, and that comprehensive validation is a large enough task that it falls outside the scope of the present model.
Finally, expandability and limitations of the NP framework are discussed.

\section{Theory}\label{sec:theory}

This section presents a constitutive model of the plastic response of a microstructural network resulting from the shear coupled motion of its boundaries, referred to here, and subsequently, as ``Grain Boundary Network Plasticity'' or, for brevity, ``Network Plasticity'' (NP).
The key idea of NP is to employ graph theory to capture the relationship between GB evolution and local deformation state.
The spatial arrangement of grains offers one way to characterize the microstructure as a graph, and forms the basis for the NP approach.
The use of graph theory has proven to be a valuable tool in analyzing and organizing large physical systems \cite{kochkarov2015issues, banerjee2019graph} whether time varying or static.
Note that graph theory has been employed in other ways, for instance, to abstract the transitions between thermodynamic states in mechanical systems \cite{banerjee2019graph}.
Graph theory has also enjoyed a resurgence due to the recent rise of graph neural networks \cite{kochkarov2015issues}.
In this work, however, the graph structure is used in a straightforward manner to encode connectedness between grains in microstructure.
This work builds on the theory developed by Chesser \etal\ \cite{chesser2020continuum}, which is a general reference for this section.

\subsection{The directed graph model of microstructure}

\begin{figure}
    \centering
    \includegraphics[width=.95\textwidth]{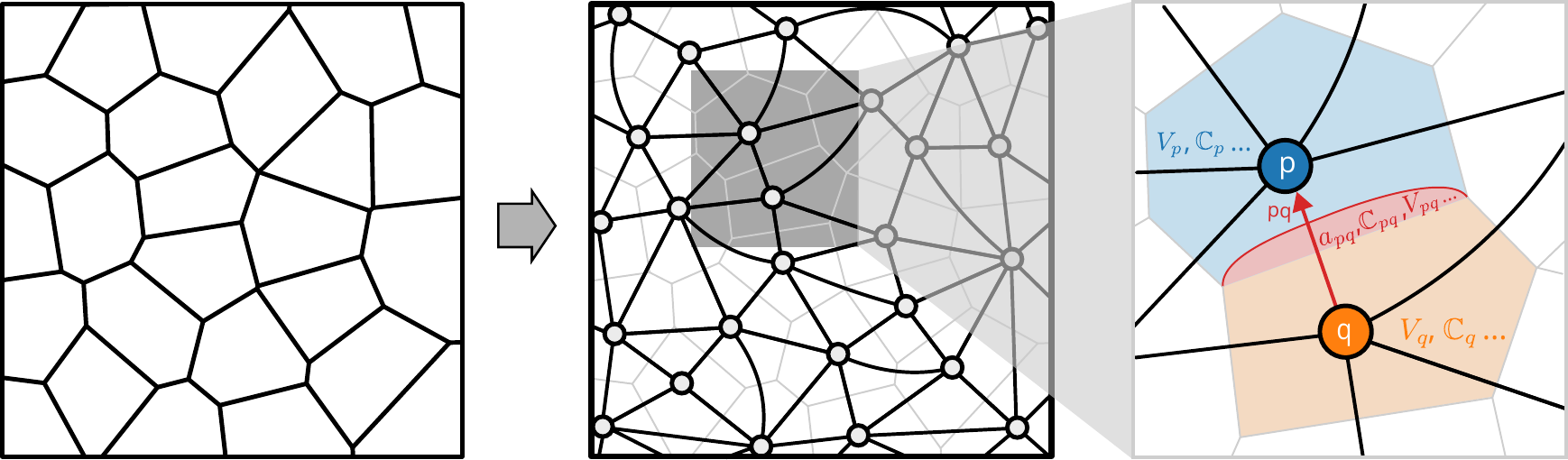}
    \caption{
      Graph representation of microstructure: A microstructure (left) is represented as a directed graph (center) with grains corresponding to nodes and boundaries to edges.
      Volumetric quantities are associated with nodes (e.g. {\color{C0} grain p} and {\color{C1} grain q} in the popout), while edge {\it and quasigrain} properties are associated with edges (e.g. {\color{C3} edge pq} undergoes motion and sweeps out quasigrain in the popout).
    }
    \label{Fig:Schmatic}
\end{figure}

Microstructure in metals can be represented as a collection of grains segregated by a network of grain boundaries, with lattice defects (such as disconnections, interstitials, vacancies) considered to be ``sub-grain'' and interfacial defects (such as disconnections or solutes) considered ``sub-boundary''.
Such a network of grains and boundaries lends itself readily to representation using the mathematical notion of a directed graph: a reduced-order abstraction that preserves connectivity, yet without resolving specific geometries of the grains or boundaries.
Note that a directed graph differs from a graph only in that its edges possess orientation.
Only directed graphs are used in this work and so ``directed graph'', ``digraph'', and simply ``graph'' are used interchangeably throughout.
One can then begin attributing microstructural quantities to the elements of a graph: volumes of grains are associated with vertices, areas of grain boundaries are associated with edges, etc.
In this section, we formalize this relationship and define the quantities of interest in the context of graph elements (\cref{Fig:Schmatic}).

\def\graph{\mathcal{G}}

We begin by defining the directed microstructural graph (``the graph'') as $\graph = (\mathbb{V},\mathbb{E})$, a collection of vertices and edges. 
In two dimensions, the graph is planar.
Vertices $\mathbb{V}\subset\mathbb{Z}$ are integer indexed, and edges $\mathbb{E}\subset\mathbb{V}\times\mathbb{V}$ are represented as tuples of vertices.
The convention is adopted that $(p,q)=pq\in\mathbb{V}$ corresponds to an edge pointing from $q$ to $p$ (i.e. $p$ is the head, $q$ is the tail).
The present notation naturally allows $\graph$ to have up to two edges, $(p,q)$ and $(q,p)$ connecting the same pair of vertices, making $\graph$ a multigraph.
Additional edges may be included if needed, as long as vertices are individually indexed.
However, $\graph$ contains no self-loops ($p,p$).
By construction, $\graph$ is always connected (all vertices are attached to all other vertices by one or more edges) but usually not strongly connected, since not every grain shares a boundary with every other grain.
In this work, $\graph$ is not considered time-varying, which means that topological transitions (merging of grains, nucleation of new grains) are not considered that involve the creation of any boundaries that were not already present.
In cases where substantial microstructure evolution occurs, transitions in the microstructure may be accounted for by evolving the structure of the graph itself.

It is essential to clarify the notation that is adopted in this work, due to the extensive use of the graph structure notation that sometimes conflicts with standard notation in mechanics.
Quantities that are subscripted with a single index (e.g. $\A_p$) shall be tacitly assumed to be vertex/grain quantities (in this case, grain $p$).
Those that are subscripted with a double index (e.g. $h_{pq}$) shall be edge/boundary quantities (in this case, between grain $p$ and grain $q$ and oriented towards $p$.)
There is no correlation between the number of indices on a variable and the character of the variable; a quantity may be a scalar, vector or tensor regardless of whether it resides at a vertex or an edge.
To avoid confusion with commonly used index notation, the indices $p,q,r,s,m,n$ will be used instead of $i,j,k$.
(We clarify that $\F_p$ refers to the deformation gradient associated with grain $p$ - not to plastic deformation.) 
Invariant notation is used to indicate vector/tensor operations: neither indicial notation nor the summation convention is used in this work.
Summations in this work are typically over vertices $p\in\verts$ and edges $pq\in\edges$ of the graph and are always written explicitly.

\begin{table}[]
    \centering\small
    \begin{tabularx}{\linewidth}{lllXX}
    \toprule 
    {\bf Symbol} & {\bf Sym} & {\bf Type} & {\bf Description} & {\bf Location/Type} \\
    \midrule
    $\vol_{p}$ &  & $\mathbb{R}$ &  Grain volume & Vertex/Grain \\
    $\F_p$ &  & $GL(3)$ & Deformation gradient & Vertex/Grain\\
    $\F^0_p$ &  & $GL(3)$ & Eigendeformation & Vertex/Grain\\
    $\P_p$ &  & $\mathbb{R}^{3\times3}$ & Piola-Kirchhoff Stress & Vertex/Grain\\
    $\mathbb{C}_{p}$ &  & $\mathbb{R}^{3\times3\times3\times3}$ &  Elastic modulus & Vertex/Grain \\
    $\A_p$ &  & $\mathbb{R}$ & Helmholtz free energy & Vertex/Grain \\
    \midrule
    $\vol_{pq}$ & $=-\vol_{qp}$ & $\mathbb{R}$ &  Swept volume & Edge/Quasigrain \\
    $\F_{pq}$ & $=\F_{qp}$ & $GL(3)$ & GB deformation gradient & Edge/Quasigrain\\
    $\F^0_{pq}$ & $=\F^0_{qp}$ & $GL(3)$ & GB eigendeformation & Edge/Quasigrain\\
    $\mathbb{C}_{pq}$ & $=\mathbb{C}_{qp}$ &  $\mathbb{R}^{3\times3\times3\times3}$ &  Elastic modulus & Edge/Quasigrain\\
    $\A_{pq}$ & $=\A_{qp}$ & $\mathbb{R}$ & Helmholtz free energy & Edge/Quasigrain \\
    \midrule
    $\area_{pq}$ & $=\area_{qp}$ & $\mathbb{R}$ &  GB area & Edge/Boundary \\
    $\nor_{pq}$ & $=-\nor_{qp}$ & $\mathbb{R}^3$ &  GB normal & Edge/Boundary \\
    $\areavec_{pq}$ & $=-\areavec_{qp}$ & $\mathbb{R}^3$ &  Grain area vector & Edge/Boundary \\
    $h_{pq}$ & $=-h_{qp}$ & $\mathbb{R}$ & GB displacement & Edge/Boundary \\
    $\Fgb_{pq}$ & $=-\Fgb_{qp}$ & Rank-1 & GB shear tensor & Edge/Boundary\\
    $\phi^*_{0pq}$ & $=\phi^*_{0qp}$ & $\mathbb{R}$ & GB dissipation energy & Edge/Boundary \\
    $\phi^*_{1pq}$ & $=\phi^*_{1qp}$ & $\mathbb{R}$ & GB inverse mobility & Edge/Boundary \\
    $\phi^*_{hpq}$ & $=\phi^*_{hqp}$ & $\mathbb{R}$ & GB hardening potential & Edge/Boundary \\
    $\kappa^*_{hpq}$ & $=\kappa^*_{hqp}$ & $\mathbb{R}$ & GB hardening onset & Edge/Boundary \\
    \bottomrule
  \end{tabularx}
  \caption{List of nomenclature used in this work. }
  \label{tab:nomenclature}
\end{table}

In the graph representation of microstructure, graph vertices correspond to single grains.
Each vertex and edge have distinct properties to describe the associated section of microstructure.
For each vertex $p$ there is associated a volume $V_p$, a deformation gradient $\F_p$, a stress tensor $\P_p$, a modulus tensor $\mathbb{C}_p$, and an eigenstrain tensor $\F^0_p$, all of which are taken to be constant throughout the grain.
We employ special linearized elasticity, which is equivalent to traditional linear elasticity except that the strain tensor is explicitly not symmetrized in order to account for grain boundary orientation.
Therefore, large-deformation mechanics quantities, such as the deformation gradient $\F$ and the Piola-Kirchhoff stress tensor $\P$ are used throughout this work.
Each grain is ascribed an eigenstrain tensor $\F^0_p$.
In this work, $\F^0_p=\I$, which is to say that there is no inelastic deformation within the grain.
The eigendeformation $\F^0_p$ may be evolved in order to account for thermal expansion or crystal plasticity.

As grains correspond to graph vertices, so grain boundaries correspond to graph edges.
To every edge there is attached an area vector $\areavec_{pq} = a_{pq}\nor_{pq}$ denoting the oriented area of the GB, with $\boldsymbol{n}_{pq}$ the oriented normal vector of the grain boundary interface corresponding to the orientation of the graph (i.e. $\nor$ points to grain $p$).
The edge is equipped with a measure of the GB displacement value $h_{pq}$, corresponding to the displacement of the boundary into grain $p$.
We note that many of the edge values possess antisymmetric, i.e. $\areavec_{pq}=-\areavec_{qp}$, $h_{pq}=-h_{qp}$.

Grain boundaries in this context move to consume one grain and expand another.
Grain consumption is captured in a simplistic way by the flow of grain volumes $V_p$, along the edges.
When no shear coupling occurs, this is sufficient.
However, to capture shear coupling makes it necessary to track the portions of grains that have experienced boundary motion, i.e., grain through which boundary has been swept.
This is because these portions of the microstructure contain eigendeformation that is distinct from that of the grain to which they have been converted.
Consequently, though the crystallographic structure of the expanded region is (usually) identical, the eigenstrains are different.
Therefore, we introduce the notion of a {\it quasigrain} that, like grain boundaries, are associated with graph edges.
The quasigrain properties associated with each edge generally follow those associated with grains: swept volume $V_{pq}$, quasigrain eigenstrain $\F^0_{pq} = \I + \Fgb_{pq}$, deformation gradient $\F_{pq}$, stress tensor $\P_{pq}$, and so on.
The material model associated with the quasigrain, $\A_{pq}$, is determined in a binary fashion based on the neighboring models $\A_p$, $\A_q$, and the sign of the boundary's motion, i.e.,
\begin{equation}
  \label{ElasticModuli_pq}
  \A_{pq}= 
  \begin{cases}
    \A_q  & h_{pq} > h^0_{pq} \\
    \A_p & h_{pq} \leq h^0_{pq}  \\
  \end{cases}.
\end{equation}
Quasigrains are thus equipped with the same properties and degrees of freedom as regular grains and are treated as regular grains in the determination of the global stress state, despite being associated with graph edges and boundaries rather than vertices.
The quasigrain eigenstrain tensor $\Fgb_{pq}$ is of particular interest in the present application, as it encodes the shear coupling properties of the boundary.

To conclude, we have introduced a robust notation and set of kinematics for a graph-theoretic representation of microstructure, which amounts to the attaching of grain-wise quantities to vertices, quasigrain-wise quantities to edges, and boundary quantities to edges.
Because this model contains many variables, both standard and unique to the present model, all quantities are enumerated in  \cref{tab:nomenclature}.

\subsection{Kinematics}

The present model relies on a simplified notion of deformation within a microstructural network.
It is therefore assumed that the deformation mapping $\phi(\bm{x})$ is grain-wise affine; that is, that $\F_p$ is constant within grain $p\in\mathbb{V}(\mathcal{G})$, and $\F_{pq}$ is constant within quasigrain $pq\in\mathbb{E}(\mathcal{G})$.
Such a mapping is known to be admissible as long as the jumps in deformation gradient is rank-1 connected with respect to the interface normal; that is, they must be Hadamard compatible.
Here, we relax this requirement, with the understanding that incompatibility-driven deformation near each boundary will not be accounted for.
Compatibility is imposed, in a weak sense only, with respect to the net average deformation experienced by the entire microstructure:
\begin{align}\label{eq:weakcompatibility}
  \F_{\text{total}} \overset{!}{=} \frac{1}{V}\Big(\sum_{p\in\mathbb{V}(\mathcal{G})}V_p\F_p + \sum_{pq\in\mathbb{E}(\mathcal{G})}|V_{pq}|\,\F_{pq}\Big),
\end{align}
where $V$ is the unchanging total volume of all grains and quasigrains. 
(Recall that $V_{pq}$ signifies quasigrain volume and is a signed quantity; $|\cdot|$ indicates the absolute value.)
The effect of this incompatibility may be accommodated in a reduced sense by modification of the elastic modulus and (as will be seen later) though the use of hardening potentials.
In the context of material point modeling, $\F_{\text{total}}$ is the state of local deformation, which must be accommodated grain-wise in an average sense.
The enforcement of \cref{eq:weakcompatibility} through Lagrange multiplies will be discussed in greater detail in \cref{Sec:Mechanical_equilibrium}.

\subsection{Shear coupling}

\begin{figure}
    \includegraphics[width=\linewidth]{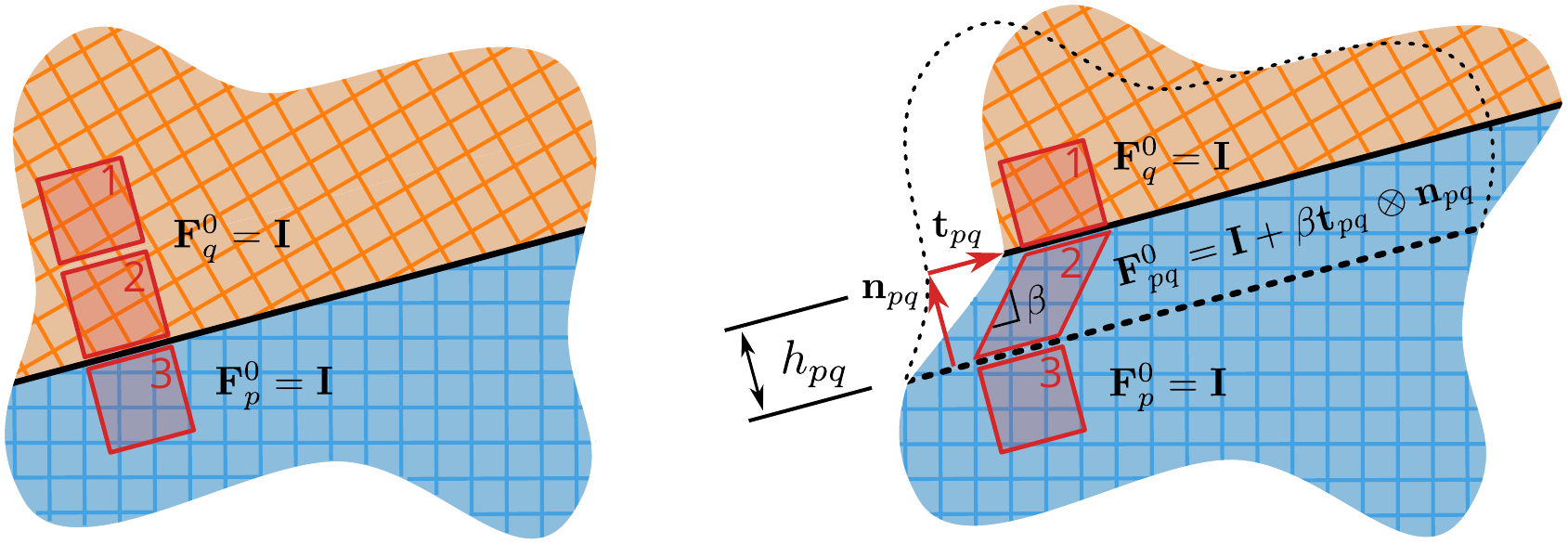}
    \caption{
    Schematic depiction of shear coupling in a generalized setting.
        Grain $p$ (gold) is converted to grain $q$ (blue) through the positive, shear-coupled motion of boundary $pq$.
        The red boxes 1-3 are fiducial markers illustrating the deformation process.
        Boxes 1 and 3 do not experience shear, and therefore have no change in eigendeformation.
        Box 2, which is within the region swept by boundary $pq$ is transformed from grain q to grain p.
        After deformation, this region posesses the same elastic properties as grain $p$; however, it has a different eigendeformation that must be tracked independently.
    This is accomplished through the use of quasigrains.
    }
    \label{fig:shearcoupling}
\end{figure}

This work follows \cite{chesser2020continuum} in the treatment of grain boundary shear coupling as an eigenstrain evolution, in which the eigenstrain of the shear coupled region is determined by the shear coupling tensor, which in turn is determined by the crystallographically-informed shear coupling factor.
For each edge/boundary $pq\in\mathbb{E}(\mathcal{G})$, there is assigned a rank-1 shear coupling matrix $\Fgb_{pq}$ that corresponds to the deformation gradient jump induced by grain boundary shear.
In fact, multiple $\Fgb_{pq}$ may be stored with their associated dissipative properties; however, here we consider a single mode only.
Shear coupling is therefore modeled in a straightforward manner by taking the eigenstrain of each quasigrain to be
\begin{align}
  \F^0_{pq} = \mathbf{I} + \operatorname{sgn}(h_{pq}) \,\Fgb_{pq}.
\end{align}
For a symmetric tilt grain boundary undergoing pure shear coupling, the jump in deformation across the boundary $pq$ is given by
\begin{align}
  \Fgb_{pq} = \bm{b}_{pq}\otimes \bm{h}_{pq} = \beta_{pq}\,\bm{t}_{pq}\otimes\bm{n}_{pq}
\end{align}
where $\bm{h}_{pq}=\bm{n}_{pq}/h_{pq}$, $h_{pq}$ is the disconnection step height, $\bm{b}_{pq}\perp\bm{n}_{pq}$ is the corresponding Burgers vector, $\bm{t}_{pq}$ is a vector tangent to the boundary, and $\beta_{pq}=|\bm{b}_{pq}|/h_{pq}$ is the shear coupling factor.
(This process is illustrated in \cref{fig:shearcoupling}.)
$\Fgb_{pq}$ is a rank-1 tensor indicating a deformation jump that is isochoric and Hadamard-compatible with respect to the boundary plane normal $\bm{n}_{pq}$.
The restriction $\bm{b}_{pq}\perp\bm{n}_{pq}$ indicates that only disconnection glide is considered; disconnection climb can be captured by using non-orthogonal vectors but is not explored in this work.
The set of possible coupling factors $\beta$ are limited to those that preserve crystallographic symmetry and can be well-approximated for symmetric tilt grain boundary (STGB).
Shear coupling factors for such boundaries have been thoroughly investigated, and so can be assigned based on the crystallography of the neighboring grains.
However, many boundaries in microstructure are asymmetric tilt grain boundaries (ATGB) which differ from STGB via the boundary orientation.
Determination of shear coupling factors for ATGBs is much less clear and is the subject of ongoing work \cite{trautt2012coupled,admal2022interface,joshi2022finite}.
In this work a simple modification rule is used to approximate ATGB shear coupling as a composition of rotations and STGB shear:
\begin{align}
\label{ATGBshearcoupling}
  \Fgb_{pq}
  = \bm{\mathrm{R}}_{pq}\mathrm{\Delta}\bm{\mathrm{F}}^{STGB}_{pq}\bm{\mathrm{R}}^T_{pq}
  = (\mathbf{R}_{pq}\bm{b}_{pq}^{STGB})\otimes(\mathbf{R}_{pq}\bm{h}_{pq}^{STGB})
\end{align}
where $\bm{\mathrm{R}}\in\operatorname{SO}(3)$ is a rotation corresponding to the inclination of the boundary plane away from the STGB orientation.
This construction follows the assumption that ATGB grains thus shear couple according to the STGB coupling factor, with the resultant incompatibility generating a substantial back-stress.
To some extent, this is consistent with some atomistic observations of ATGB migration \cite{trautt2012coupled}, but we leave the full exploration of ATGB shear coupling factor approximation to future work.

\subsection{Mechanical equilibrium} \label{Sec:Mechanical_equilibrium}

The computation of equilibrium grain deformations $\F_{p\in\mathcal{V}(\graph)}$ and quasigrain deformations $\F_{pq\in\mathbb{E}(\mathcal{G})}$ follows from the principle of minimum potential energy, constrained by the weak compatibility requirement.
As each grain and quasigrain possesses a Helmholtz free energy, the total Helmholtz free energy of the entire system is given by
\begin{equation}\label{Total_Helmholtz_Free_Energy}
\A(\{\F_{p\in\mathbb{V}(\mathcal{G})}\},\{\F_{pq\in\mathbb{E}(\mathcal{G})}\}) = \frac{1}{V}\Big(\sum_{p\in\mathbb{V}(\graph)} V_pA_p(\F_p) + \sum_{pq\in\mathbb{E}(\graph)} |V_{pq}|A_{pq}(\F_{pq})\Big).
\end{equation}
Consequently, the equilibrium values for the grain- and quasigrain-wise deformations follow as the solution of the variational free energy minimization problem,
\begin{align}\label{eq:minimum_helmholtz}
    \A(\F) = \inf_{\{\F_p\},\{\F_{gb}\}} \A  \ \ \ \text{subject to}  \ \ \ \sum_{p\in\mathbb{V}(\graph)}V_p\,\F_p + \sum_{pq\in\mathbb{E}(\graph)} |V_{pq}|\,\F_{pq} = V\F.
\end{align}
The choice of grain-wise Helmholz energies is arbitrary up to the usual material model constraints (material frame indifference, positive definiteness, etc), and may feature linear elastic, nonlinear elastic, or inelastic behavior as required.
Recall that the choice of quasigrain-wise material models is dictated by \cref{ElasticModuli_pq}.
In this work, we invoke special linearized elasticity, i.e. elasticity linearized about each grain's eigenstrain,
\begin{gather}\label{Grain_Helmholtz_Free_Energy}
\begin{array}{rl}
     \A_p(\F_p) &= \frac{1}{2}(\F_p-\F^0_p):\mathbb{C}_p(\F_p-\F^0_p) + u_{p}(\F_p,\boldsymbol{q}_{p},T_{p}) \ \ \ p\in\mathbb{V}(\mathcal{G})\\
     \A_{pq}(\F_{pq}) &= \frac{1}{2}(\F_p-\F^0_{pq}):\mathbb{C}_{pq}(\F_{pq}-\F^0_{pq}) + u_{pq}(\F_{pq},\boldsymbol{q}_{pq},T_{pq}) \ \ \ pq \in \mathbb{E}(\mathcal{G}),
\end{array}
\end{gather}
where the $:$ product on two tensors is the Frobenius inner product, $\bm{\mathrm{A}}:\bm{\mathrm{B}}=\operatorname{tr}(\bm{\mathrm{A}}^T\bm{\mathrm{B}})$, and $u_{p},u_{pq}$ are additional potentials (for instance, a synthetic driving force), that may depend on other internal variables $\mathbf{q}$ or temperature $T$.
The potential $u$ can also represent a body potential that encourages or discourages grain growth based on ambient conditions, such as is observed in grain growth of thin films \cite{thompson1996stress,estrin2001grain}. In this work, however, $u$ is considered to be zero, and no additional internal variables are considered.
In this work, we regard the eigenstrain for each grain to be $\F^0_p=\mathbf{I}$; however, this may be evolved using a plasticity flow rule to account for other dislocation-based mechanisms of deformation.
The quasigrain eigenstain contains shear-coupling deformation, and will be discussed in the subsequent section.

The elastic modulus tensors $\mathbb{C}$ contain the elastic modulus as linearized about the eigendeformation $\F^0$, and contain the major and minor symmetries associated with linear elasticity.
Note, however, that the kinematic variable is the deformation gradient, not the usual symmetrized strain tensor.
This is necessary to properly account for the directionality of deformation during shear coupling, but does not otherwise affect the elastic solution.
The average Piola Kirchoff stress tensor is determined by the derivative of \cref{Total_Helmholtz_Free_Energy}:
\begin{equation}
  \label{P0eq}
  \P = \frac{1}{V}\sum_{p\in\mathbb{V}(\graph)} V_p\P_p(\F_p-\bm{I}) + \frac{1}{V}\sum_{pq\in\mathbb{E}(\graph)} |V_{pq}|\P_{pq}(\F_{pq}-\F^0_{pq}).
\end{equation}
The solution of \cref{eq:minimum_helmholtz} for optimal grain-wise and quasigrain-wise deformation gradients is derived in \ref{sec:mechanical_equilibrium} and can be expressed in closed-form as,
\begin{gather}
    \F^*_p
    =
    \I + 
    \mathbb{C}^{-1}_p
    \langle\mathbb{C}\rangle\Bigg(\F-\I
    -
    \frac{1}{V}\sum_{rs\in\edges}|V_{rs}|\Fgb_{rs}\Bigg) \ \ \ \ \forall p \in \verts\label{eq:fp_optimal}\\
    \F^*_{pq}
    =
    \I + \Fgb_{pq} +  
    \mathbb{C}^{-1}_{pq}
    \langle\mathbb{C}\rangle\Bigg(\F-\I
    -
    \frac{1}{V}\sum_{rs\in\edges}|V_{rs}|\Fgb_{rs}\Bigg) \ \ \ \ \forall pq \in \edges, \label{eq:fpq_optimal}
\end{gather}
where the asterisk indicates optimality (that is, variable with an asterisk solves the corresponding optimization problem - in this case, $\mathbf{F}^*_p,\mathbf{F}^*_{pq}$ are the solution to \cref{Total_Helmholtz_Free_Energy}) here and throughout this work.
The solution for the grain and quasigrain deformation gradients is given in terms of the network averaged elastic modulus tensor $\langle\mathbb{C}\rangle$, which is 
\begin{align}\label{eq:average_c}
    \langle\mathbb{C}\rangle = \Bigg[\frac{1}{V}\Big(\sum_{r\in\verts}V_r\mathbb{C}_r^{-1}
    + 
    \sum_{rs\in\edges}|V_{rs}|\mathbb{C}_{rs}^{-1}\Big)\Bigg]^{-1}.
\end{align}
The corresponding stress, which is shown to be constant throughout all of the grains, is determined to be
\begin{align}
  \P = \P_p(\F^*_p) = \P_{pq}(\F^*_{pq}) = \langle\mathbb{C}\rangle \Bigg((\F-\I)
  -
  \frac{1}{V}\sum_{rs\in\edges}V_{rs}\Fgb_{rs}\Bigg) \ \ \ \ \forall p\in\verts, \forall pq\in\edges,
\end{align}
concluding the mechanical equilibirium calculation.
  The result that stress is constant in all grains is aligned with the assumption made by Sachs \cite{sachs1928ableitung}, making this model analagous to a Sachs-type treatment.
  (The alternative is a Taylor-type treatment, which assumes consistent strain and variable stress).
The closed-form solution to \cref{eq:minimum_helmholtz}, presented here, relies on the use of special linearized elasticity.
If a nonlinear material model is used, then it must be solved numerically.

In some cases, it may be preferable to prescribe the average stress tensor $\P$ rather than the deformation gradient $\F$.
If the stress tensor is prescribed, the minimization problem becomes
\begin{align}
    \F^*_{p\in\verts}, \F^*_{pq\in\edges} = 
    \underset{\F^*_{p\in\verts}, \F^*_{pq\in\edges}}{\arg \sup}\Bigg[\P:\F - \A(\F_{p\in\verts},\F_{pq\in\edges})\Bigg],
\end{align}
where $\F$ is unknown, relaxing the requirement of weak compatibility.
Stationarity for the deformation gradients implies
\begin{align}
    \F^*_p &= \I + \mathbb{C}_{p}^{-1}\P
    &
    \F^*_{pq} &= \I + \sgn(h_{pq})\Fgb_{pq} + \mathbb{C}_{pq}^{-1}\P,
\end{align}
and the total Gibbs' energy becomes
\begin{align}
    G(\P) &= \frac{1}{V}\sum_{pq\in\edges}V_{pq}\P:\Fgb_{pq} 
    +
    \frac{1}{2}\P:\langle\mathbb{C}\rangle^{-1}\P
    + \P:\I.
\end{align}
It is then straightforward to show that $\G(\P)$ is the Legendre transform of $\A(\F)$, as expected.

\subsection{Principle of minimum dissipation potential} \label{Sec:MinDissPot}

In the network model of microstructure evolution, the grain boundary positions $h_{pq\in\mathbb{E}(\graph)}$ represent the evolution of microstructure along the edges of the graph.
These variables are like internal variables in plasticity, such as accumulated slip on a glide plane.
Therefore, the objective is to determine an evolution law for these boundary position variables.

The concept of modeling dissipative systems variationally has its roots in the original work of Onsager \cite{onsager1931reciprocal} as an extension to Rayliegh's least dissipation of energy \cite{Rayleigh}.
The general theme is that the irreversible loss of energy to heat can be represented as a dissipation potential.
Energy balance dictates that any change in free energy must correspond to a change in dissipated energy; from this, the principle of minimum dissipation potential may be constructed.
Variations of this type of method have been successfully employed towards the modeling of various systems, particularly viscoplastic materials \cite{chaouki2016viscoplastic,mcbride2018dissipation}.
It has recently been employed to study the migration of grain boundaries, both at the continuum level \cite{chesser2020continuum,wei2020grain} and the mesoscale \cite{runnels2020phase,gokuli2021multiphase}.
The advantage of this framework is derived from its variational structure, which is beneficial to the formulation and solution of problems, especially when many degrees of freedom are present (as in the present case).
As always, with models of this type, it should be emphasized that the ``principle'' of minimum dissipation potential is inherently heuristic, and that caution should always be taken when extrapolating the idea of minimization to deriving other physical insights.

The principle of minimum dissipation potential for a dissipative system is
\begin{align}
  \dot{\bm{q}} = \underset{\dot{\bm{q}}}{\operatorname{arg}\inf}\Big[\frac{\partial}{\partial t}\mathrm{W}(\dot{\bm{q}}) + \phi^*(\dot{\bm{q}})\Big],
\end{align}
where $\mathrm{W}$ is a generalized free energy whose time derivative represents the dissipation of free energy, $\dot{\bm{q}}$ is a vector of internal variables, and $\phi^*$ is a dissipation potential.
In the present context, $\dot{\bm{q}}$ are the interface velocities.
The kinetic equation, then is given by
\begin{equation}
\label{hDotMin}
\{\Dot{h}\} = \mathop{\mbox{arg inf}}_{\{\Dot{h}\}} \Bigg[\frac{\partial}{\partial t}\A(\F) + \sum_{pq\in\mathbb{E}(\graph)} a_{pq}(\phi_{pq}^*(\Dot{h}_{pq}) )\Bigg].
\end{equation}
Stationarity for $h_{pq}$ yields the following expression
\begin{align}
0 &\in 
    \mathrm{DA}_{pq} +  \frac{\partial\phi_{pq}^*}{\partial\Dot{h}_{pq}},
  &
    \mathrm{DA}_{pq} = \frac{1}{a_{pq}}\frac{\partial A}{\partial h_{pq}}
\end{align}
where the quantity $\mathrm{DA}_{pq}$ represents the thermodynamic driving force conjugate to the motion of boundary $pq$, per unit grain boundary area. 
Note that the subderivative is taken of $\phi^*_{pq}$, which may be a set of finite measure if $\phi^*$ is not smooth.
Therefore, stationarity requires that zero be contained within the derivative, not necessarily equal to it.
The dissipation potential $\phi^*$ admits a fairly general form \cite{onsager1931reciprocal}, but in this work, the following is used:
\begin{equation}
  \phi^*_{pq}(\dot{h}_{pq}) = \phi^*_{pq0}|\dot{h}_{pq}|+\frac{1}{2}\phi^*_{pq1}|\dot{h}_{pq}|^2,
\end{equation}
where $\phi_{pq0}^*$ is a rate-independent yield constant, and $\phi_{pq1}^*$ is a rate-dependent time constant.
  The yield constant, also referred to as the ``dissipation energy'' corresponds to the critical intensive driving force necessary to initiate motion of the boundary.
  It can be calculated for a bicrystal in molecular dynamics by integrating a stress-strain loading curve or, alternately, by measuring the synthetic driving force required to initiate motion \cite{chesser2020continuum}.
  The time constant is the inverse of the well-known and recorded grain boundary mobility.
The use of the absolute value induces, crucially, a lack of smoothness in the dissipation potential that must be considered in the minimization process.
Substituting this form into the stationarity condition produces the following explicit nonlinear kinetic relation for $\dot{h}_{pq}$:
\begin{align}\label{eq:kineticequation}
    \dot{h}_{pq} = -\frac{1}{\phi^*_{pq1}}
    \begin{cases}
        \mathrm{DA}_{pq} - \phi^*_{pq0} & \mathrm{DA}_{pq} > +\phi^*_{pq0} \\
        \mathrm{DA}_{pq} + \phi^*_{pq0} & \mathrm{DA}_{pq} < -\phi^*_{pq0} \\
        0 & \mathrm{else}
    \end{cases}.
\end{align}
\Cref{eq:kineticequation} can be explicitly or implicitly integrated in time to obtain the evolution of the boundary locations.
However, the computation of $\mathrm{DA}_{pq}$ requires the evaluation of several nontrivial derivatives of volume and equilibrium free energy with respect to interface location. 

In the case of an applied load ($\P$), the Gibbs energy is used instead of the Helmholtz energy. 
So, the kinetic equation under an applied load is
\begin{align}\label{eq:kineticequation_P}
    \dot{h}_{pq} = -\frac{1}{\phi^*_{pq1}}
    \begin{cases}
        \mathrm{DG}_{pq} - \phi^*_{pq0} & \mathrm{DG}_{pq} > +\phi^*_{pq0} \\
        \mathrm{DG}_{pq} + \phi^*_{pq0} & \mathrm{DG}_{pq} < -\phi^*_{pq0} \\
        0 & \mathrm{else}
    \end{cases},
\end{align}
where $\mathrm{DG}$ is defined in the same way as $\mathrm{DA}$.
Explicit expressions for both are defined in the following sections.

\subsection{Volume flow calculation}

In the context of a graph representation, microstructure evolution is represented as flow of volume along edges between vertex and edge quantities.
As discussed previously, many of the kinematic restrictions common to models of this type (e.g. Hadamard compatibility), are relaxed in order to avoid costly pointwise resolution of displacement fields.
There are two primary kinematic considerations in the construction of this model: (1) rules governing the flow of volume between edges and vertices; and (2) methods for the construction of grain boundary shear tensors.

The interface positions $h_{pq}$ govern the evolution of the microstructure following the principle of minimum dissipation.
As $h_{pq}$ evolves, the effect on the associated quasigrain is straightforward through the quasigrain volume flow relation $dV_{pq} = a_{pq}\,dh_{pq}$, which implies that
\begin{align}\label{eq:dVrs_dhpq}
  \frac{\partial V_{rs}}{\partial h_{pq}} = \begin{cases}a_{pq} & rs = pq \\ 0 &\text{else}\end{cases}.
\end{align}
The evolution of the grain-wise volumes is more complex; recalling that grain-wise volumes shrink when quasigrains grow into them, but remain the same when quasigrains grow out of them, the derivative of volume with respect to interface position becomes
\begin{align}\label{eq:dV_dh}
  \frac{\partial V_r}{\partial h_{pq}} = 
  \begin{cases}
    -a_{pq}  & r=p \text{ and } V_{pq} > 0 \\
    +a_{pq}  & r=q \text{ and } V_{pq} < 0 \\
    0 & \text{ else }
  \end{cases}.
\end{align}
Careful and precise volume updates are crucial to the accurate evolution of the interface positions.
The derivatives \cref{eq:dVrs_dhpq,eq:dV_dh} above are primarily used for driving force calculations below.
While they can also be integrated in time to track volumes, it is more practical and accurate to perform volume updates based on the original volume distributions:
\begin{align}\label{eq:grainvolumeupdate}
  V_p = V_p^0 - \sum_{pq\in\mathbb{E}(\graph)}\max(V_{pq},0) + \sum_{qp\in\mathbb{E}(\graph)}\min(V_{qp},0),
\end{align}
where $V_p^0$ tracks the original grain-wise volume, $V_{pq}=a_{pq}h_{pq}$ the signed quasigrain volume, and the summations indicate the contributions from the various neighboring quasigrains.

\subsection{Driving force for a prescribed deformation}

  Of particular interest is the case in which a prescribed deformation, $\mathbf{F}$ is applied to the graph microstructure.
  As a consititutive model, NP is ideal for use in the context of large-scale continuum simulations (finite element analysis) to bridge between local microstructure stress state and continuum-level boundary conditions.
Such a model typically requires a closed-form expression for the Helmholz free energy as a function of deformation gradient, $\mathrm{A}(\mathbf{F})$, first and second derivatives thereof, as well as a mechanism for evolving internal variables through time.

  The key quantity is the GB-wise thermodynamic driving force, $\mathrm{DA}_{pq}$, the calculation of which leads to the grain- and quaisgrain-wise deformation gradients $\mathbf{F}^*_{p\in\verts},\mathbf{F}^*_{pq\in\edges}$.
  This leads to the calculation of the Helmholtz energy and its derivatives (\protect\cref{Total_Helmholtz_Free_Energy}), and the internal variable evolution through \protect\cref{eq:kineticequation}.
  To calculate $\mathrm{DA}_{pq}$, we begin by direct evaluation of the derivative with respect to GB position:
\begin{align}
\label{DrivingForce}
\mathrm{DA}_{pq} &= \frac{1}{V\,a_{pq}}\frac{\partial}{\partial h_{pq}}\Big( \sum_{r\in\mathbb{V}(\graph)} V_r\A_r^* + \sum_{rs\in\mathbb{E}(\graph)} |V_{rs}|\A_{rs}^*\Big) \\
                 &=    \frac{1}{V}\Bigg(
                   \sum_{r\in\mathbb{V}(\graph)}\Big(
                   \frac{\partial V_r}{\partial h_{pq}}\,\A^*_r
                   +
                   V_r\P^*_r:\frac{\partial \F^*_r}{\partial h_{pq}}\Big)
                   +
                   \sum_{rs\in\mathbb{E}(\graph)}\Big(\frac{\partial|V_{rs}|}{\partial h_{pq}} \A^*_{pq}
                   +
                   |V_{rs}|\P^*_{rs}:\frac{\partial \F^*_{rs}}{\partial h_{pq}}
                   \Big)\Bigg),
\end{align}
where we recall that the asterisk implies optimality, i.e. $\A_p^*=\A_p(\F^*_p)$, and so on.
The expression for driving force may be simplified by noting that $A^*_{pq}=A^*_q$ when $h_{pq}>0$, and $A^*_{pq}=A^*_p$ when $h_{pq}<0$:
\begin{gather}
  \sum_{r\in\verts}\frac{\partial V_r}{\partial h_{pq}}A^*_r + \sum_{rs \in \edges} \frac{\partial |V_{rs}|}{h_{pq}}A^*_{pq}
  =
  a_{pq}(A^*_q - A^*_p).
\end{gather}
Further simplification requires the differentiation of the equilibrium grain-wise and quasigrain-wise deformation gradients with respect to interface position.
Since a closed-form solution exists, it is possible to calculate the derivatives explicitly.
The calculation is presented in \ref{sec:derivation_driving_force_terms}, and results in
\begin{align}
    \frac{\partial\F^*_r}{\partial h_{pq}}
  &=
    -\frac{a_{pq}}{V}
    \mathbb{C}^{-1}_r
    \langle\mathbb{C}\rangle
    \Big[
    \Fgb_{pq} + \F^*_q - \F^*_p
    \Big] 
  &
    \frac{\partial\F^*_{rs}}{\partial h_{pq}}
    =
    -\frac{a_{pq}}{V}
    \mathbb{C}^{-1}_{rs}
  &\langle\mathbb{C}\rangle\Big[
    \Fgb_{pq} + \F^*_q - \F^*_p\Big]. \label{eq:dF_dh}
\end{align}
Substituting these into the expression for $\mathrm{DA}_{pq}$ yields 
\begin{align}
  \mathrm{DA}_{pq} =    \frac{1}{V}\Bigg(
  (A^*_q - A^*_p)
  -
  \P^*_{pq}
  :
  \underbrace{\Bigg[
  \frac{1}{V}
  \sum_{r\in\mathbb{V}(\graph)}\Big(
  V_r
  \mathbb{C}^{-1}_r
  \Big)
  +
  \frac{1}{V}
  \sum_{rs\in\mathbb{E}(\graph)}\Big(
  |V_{rs}|
  \mathbb{C}^{-1}_{rs}
  \Big)
  \Bigg]}_{\langle\mathbb{C}\rangle^{-1}}
  \langle\mathbb{C}\rangle
  (\Fgb_{pq} + \F^*_q-\F^*_p)
  \Bigg),
\end{align}
the final expression for the driving force under a prescribed deformation, $\F$, for grain boundary $pq$, is
\begin{align}
  \mathrm{DA}_{pq} &=    \frac{1}{V}\Bigg(
  (\A^*_p - \A^*_q)
  - \P^*:\Fgb_{pq}
  \Bigg).
\end{align}

\subsection{Driving force for a prescribed load}

The derivation for $\mathrm{DG}$ is substantially simpler than that for $\mathrm{DA}$, because the equilibrium grain/quasigrain-wise deformation gradients do not change with respect to interface positions.
  While the calculation of $\mathrm{DG}$ is less cumbersome, however, the Gibbs free energy is less useful for continuum-level calculations.
  The process is included here for completeness, and also because calculation of $\mathrm{DG}$ is useful in the calculation of yield surfaces.
Calculating the derivative explicitly (and using \cref{average_cinv_dhpq_calc}) yields,
\begin{align}
    \mathrm{DG}_{pq} = \frac{1}{a_{pq}}\frac{\partial \G(\P)}{\partial h_{pq} }
  = \frac{1}{V}\mathbf{P}:\Fgb_{pq}
   + \frac{1}{2V}\P:(\mathbb{C}^{-1}_q-\mathbb{C}^{-1}_p)\P.
\end{align}
The Gibbs driving force is useful for the calculation of yield surfaces.
The non-hardened yield surface is given by
\begin{align}
  \partial\Bigg[\bigcap_{pq\in\edges}\operatorname{supp}\Big(|\mathrm{DG}_{pq}(\P)| < \phi^*_{pq0}\Big)\Bigg]
\end{align}
where $\partial$ denotes the boundary operator and $\operatorname{supp}$ denotes the support.
The calculation of a non-hardened yield surface does not require an equilibrium solve.
To determine the evolution of the yield surfaced, the contribution from hardening potentials must be included with $\mathrm{DG}$. 

\subsection{Mesoscale geometric hardening for the approximation of non-planar polycrystal effects}

\begin{figure}
    \centering
    \includegraphics[width=\textwidth]{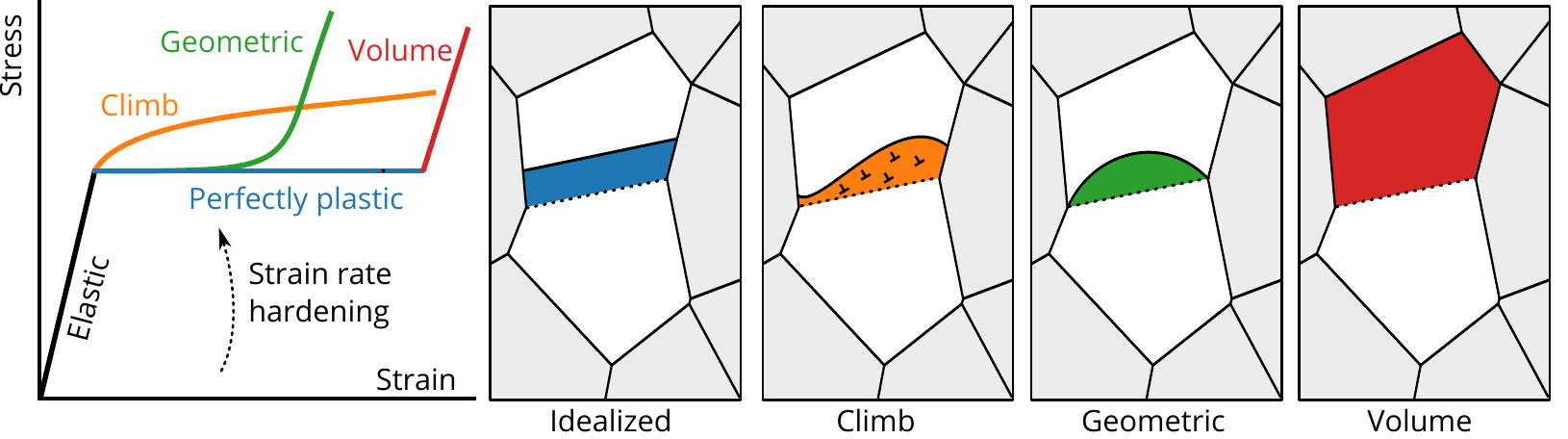}
    \caption{
      Hardening mechanisms during GB migration.
      {\color{C0} Idealized}: No hardening, boundary couples freely under shear corresponding to STGB orientation;
      {\color{C1} Climb}: Boundary interaction with dislocations, emission of dislocations, tilt of boundary to account for incompatibility;
      {\color{C2} Geometric}: Frustration of boundary motion due to triple point pinning
      {\color{C3} Volume}: Entire grain consumed by boundary, motion stopped abruptly.
    }
    \label{Fig:hardening}
\end{figure}

  The present model relies upon the approximation of all grain boundaries as planar interfaces.
  In reality, boundaries typically exhibit non-planar effects at all scales, ranging from atomic-scale microfaceting to triple-point pinning.
  Such geometric effects may have a substantial effect on the migration of the boundary, and consequently, the overall microstructure evolution \cite{rabkin2005effect,salama2020role}.
  The existence of such features as facets can sometimes be estimated predictively \cite{runnels2016relaxation}, and used to interpret \cite{chesser2018understanding} or model \cite{qiu2023interface} shear-coupled migration.
  As these features exist at scales well below the resolution of NP, they must be modeled through the modification of the GB dissipation potential.

  On the other, geometric features induced by the morphology of the microstructure cannot be accounted for merely through the adjustment of the dissipation potential.
  Most microstructures present obstacles to idealized grain boundary that frustrate boundary motion and, consequently, produce hardening.
  We refer to this behavior as ``geometric hardening,'' and here propose a heuristic mechanism for capturing such behavior in a reduced order manner.
This approach draws inspiration from previous efforts, in which changes in material geometry that describe porous media densification \cite{narayanasamy2008effect} were captured as a hardening rule.
Texture evolution-dependent hardening in HCP polycrystals is another example of coupling geometric effects to plastic behavior \cite{esque2018back}.

For the present work, we simplify the range of possible geometric effects to those resulting from the pinning effect of triple points in microstructure, which have a clear impact on GB motion and engineering \cite{aramfard2014influences,wei2020grain,randle1999mechanism}.
The range of boundary migration behaviors may be categorized in four ways (\cref{Fig:hardening}).
First, the boundary may move with no pinning effect, in exactly the same way it would as if it were within a bicrystal.
Such motion corresponds to an elastic-perfectly-plastic response.
Second, boundaries whose deformation mode is not compatible with the applied deformation may move at the expense of generating back-stress and/or emission of defects resulting from dislocation climb \cite{wei2022direct,rajabzadeh2014role,admal2022interface}.
This has an immediate impact on the plastic behavior, causing immediate hardening.
We refer to this conceptually as ``climb'' hardening.
Third, the boundary may become pinned at one or both ends, resulting in deformation and curvature of the boundary \cite{aramfard2014influences}.
The degree to which the boundary is pinned influences the plastic behavior: as the boundary becomes curved, it becomes incompatible with the shear coupling deformation, frustrating the boundary motion.
Additional elastic or plastic deformation must then be generated to ensure continuity over the interface \cite{thomas2019disconnection}.
We refer to this effect as ``geometric'' hardening.
Finally, after excessive boundary motion, a grain may be completely consumed, causing an immediate cessation of plastic flow.
We refer to this effect as ``volume'' hardening. 
Realistic boundary motion generally is composed of a combination of bowing, pinning, and unpinning of the boundary, resulting in the motion of triple points \cite{wei2020grain}.
Such processes are strongly dependent on temperature and crystallography, and only become relevant after a substantial amount of deformation has occurred.
Therefore, it is assumed here that triple junctions are relatively immobile during stress-induced GB motion, which is also consistent with Armafrad \etal. \cite{aramfard2014influences}.

The model for geometric hardening follows by analogy to work hardening in other plasticity models.
To recover the effect of energy lost due to the hardening effect, a hardening potential, $\phi^h_{pq}(h_{pq})$ is added to the dissipation potential.
The exact form of the hardening potential $\phi^h_{pq}$ is immaterial to the network plasticity framework and can be chosen by theretical modeling, empirical fitting, or machine learning.
  For purposes of demonstration, we present an empirical power-law hardening, inspired by models for latent hardening \cite{ortiz1999nonconvex}, and modeled using the form:
\begin{equation}
\label{geometricHardening}
    \phi^h(h_{pq}) = \frac{\phi^h_{pq1}}{n}\left|\frac{h_{pq}}{\kappa_{pq}}\right|^n+\frac{\phi^h_{pq2}}{m}\left|\frac{h_{pq}}{\kappa_{pq}}\right|^m,
\end{equation}
with $1 < n < 2, m > 2$, $\phi^h_{1}$, and $\phi^h_{2}$ fitting parameters specific to the boundary.
(It is important to note at this stage that the form for $\phi^h_{pq}$ is general and amenable to advanced calibration using atomistic calculations; here, a simple yet effective form is used for convenience.)
The parameter $\kappa$ controls the onset of hardening and must, in general, be determined phenomenologically.

  In this work we use, for illustrative purposes, a heuristic model for geometric hardening that is based on the observation that GBs pinned by triple junctions tend to bow out \cite{maier2014theoretical}, frustrating migration and effectively causing hardening.
  Given the paucity of comparison data (either experimental or atomistic), we propose the simple approximation that hardening is proportional to the bowed-out area of the GB.
  The area of the bowed GB is assumed to follow a parabolic profile in 2D (whereas in 3D it may be expected to form a spherical cap), and the height of the parabola is taken to be the GB migration distance, $h_{pq}$.
This relationship can be analytically calculated by the integral formula
  \begin{equation}
    a^{\mathrm{eff}}_{pq} \approx \int_{-a_{pq}/2}^{a_{pq}/2}\frac{h_{pq}}{\mu}(x-a_{pq}/2)(x+a_{pq}/2)ds = 
    \frac{1}{2}\left[\frac{a_{pq}}{2}\sqrt{\frac{h_{pq}^2a^2_{pq}}{\mu^2} + 1} - \frac{\mu}{2h_{pq}}\mbox{sinh}^{-1}\left(\frac{-h_{pq}a_{pq}}{\mu}\right)\right].
  \end{equation}
  With the bow-out area, we then estimate the hardening as proportional to the ratio of the original area to the bow out area, multiplied by the minimum grain volume to create a smooth transition from geometric to volume hardening.
  The expression for $\kappa_{pq}$ is,
\begin{equation}
    \kappa_{pq} = \frac{a_{pq}(\mbox{min}(V_p,V_q) + \epsilon)}{a^{\mathrm{eff}}_{pq}f},
\end{equation}
where $f$ is a fitted correction factor, typically correcting for the out-of-plane area change, and $\epsilon$ is a small number to prevent division by zero when $V_p$ or $V_q$ vanishes.
The remaining parameters in the hardening potential are, in general, crystallography dependent; however, there is currently insufficient experimental or atomistic data to do a rigorous fitting.
Therefore, in the present work, a constant value will be used to model hardening across all boundaries.
Stagnation of boundaries can in general be fitted by the parameters $\phi^h_{2}$ and $m$, to data from other techniques; on the other hand, $\phi^h_{1}$ and $n$ are used to fit any deviations from the elastic perfectly plastic relationship may exhibit from boundary motion. 
Then the GB evolution is given by minimizing the combination of potential energy change, dissipation potential, and hardening potential with respect to interface velocities.
Note that the interface position in geometric hardening is recoverable, which means that there is no need to track accumulated damage beyond the motion of the interface itself.
This relies on the assumption that the incompatibility-accommodating strain is all elastic rather than plastic.

  In this work we have presented a basic hardening potential, but this mechanism is a venue for much more extensive model couplings than are explored here.
  For instance, the effect of dislocation pile-up near boundaries can have a profound mechanical effect \cite{joshi2017interacting,xin2016effect}.
  This may be conceivably rendered as a hardening mechanism with an associated potential, dependent upon not only the NP variables, but crystal plastic variables as well.
  Another interesting application is the inclusion of grain boundary segregation as a hardening mechanism, which is known to affect mechanical properties \cite{babicheva2016effect} as well as to be induced by mechanical loading \cite{xu2020deformation}.
  Such behavior could be approximated based on phenomenological estimates of boundary segregation or, to achieve a more thorough coupling, by modeling segregation as a flow along the graph structure that is directed by atomistically-derived grain and boundary properties.
  Modeling of these kinds of behavior are well out of scope of the present work, but represent some possibilities in model scale-up via the NP hardening approach.

\section{Examples}\label{sec:examples}
In this section, the network plasticity framework is used on several examples including matching atomistic simulations stress strain curves, determining the effect of the hardening potential, and reconstructing GB engineered microstructures from electron back-scatter diffraction (EBSD) data from literature.
The lack of available mechanical data for EBSD microstructure makes comprehensive validation complex, since {\it in situ} mechanical data are needed along with microstructure evolution measurements.
Consequently, calibration tests for NP will be presented first (dissipation energy, mobility, hardening, etc), and then NP will be applied to a selection of realistic microstructures to determine its range of mechanical responses and convergent behavior as the number of grains increases.

  For purposes of generating the results in this work, NP was tested using a python prototype script.
  Python datastructures (set, dict, etc) were used along with the NumPY \cite{harris2020array} for linear algebra operations, accelerated with Cython \cite{behnel2011cython}, to enable intuitive code without prohibitively slow performance.\protect\footnote{Code is available upon request.}
  For large-scale computation with more than a few hundered grains, or to integrate NP into a finite element framework, a compiled code (such as C++) with highly efficient data structures must be used. 

\subsection{Symmetric tilt grain boundaries}\label{sec:ex_stgb}

\begin{figure}
  \centering
  \begin{subtable}{0.5\linewidth}
    \begin{tabularx}{\linewidth}{lXXXXX}
      \toprule
      Prop.& GB3 & GB4 & GB25 & GB117 & GB40 \\
      \midrule
      $\Sigma$ & 3 & 3 & 21 & 27 & 13\\
      Tilt Ax. & \hkl[110] & \hkl[110] & \hkl[111] & \hkl[110] & \hkl[100]\\
      $\theta$ & $109.5^\circ$ & $70.5^\circ$ & $148.4^\circ$ & $42.1^\circ$ & $67.4^\circ$ \\
      \midrule
      $\phi_0$ [GPa] & 6.3 & 0.1 & 0.01 & 0.0 & 3.8\\
      $\phi_1$ [s/GPa]& 0.001 & 0.01 & 0.00025 & 0.004 & 0.001\\
      $\beta$ [1] & 1.5 & 1.4 & -0.57 & -0.77 & 1.5\\
      $\phi^h_{1}$ [Gpa/s] & - & 0.65 & - & - & - \\
      $n$ [1] & - & 1.03  & - & - & -\\
      $f$ [1] & - & 10  & - & - & -\\
      $\mu$ [1] & - & 10 & - & - & -\\
      \midrule
    \end{tabularx}
  \end{subtable}%
  \begin{subfigure}{0.5\linewidth}
    \includegraphics[height=6cm]{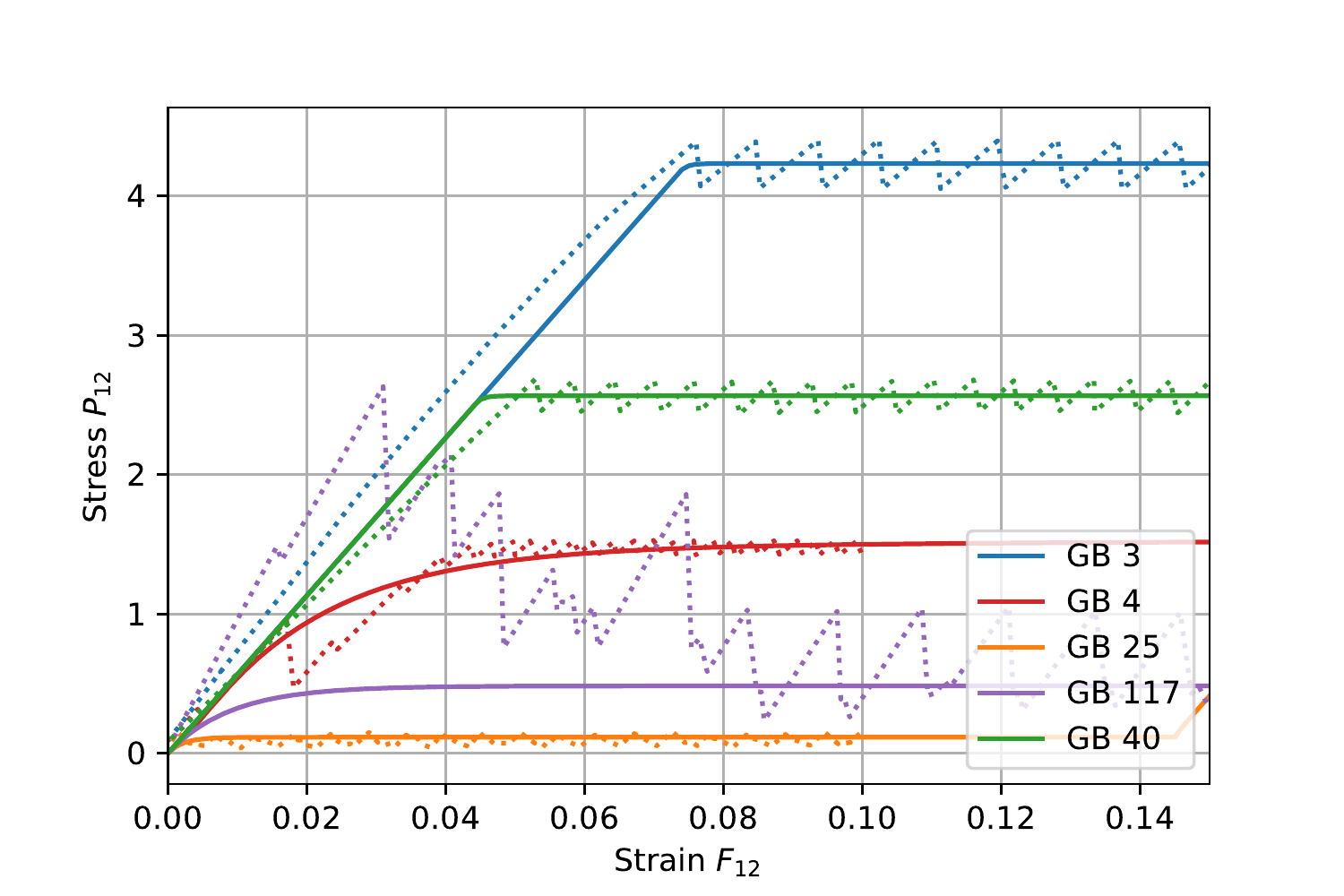}
  \end{subfigure}
  \caption{
    Network plasticity model used to simulate symmetric tilt grain boundaries from \cite{chesser2020continuum}.
    (Left) Table of parameters used for the five different boundaries.
    (Right) Comparison of stress-strain response under pure shear for aluminum ($C_{11}=108.2MPa$, $C_{12}=61.3MPa$, $C_{44}=28.3MPa$).
    Dashed lines indicate data from \cite{chesser2020continuum}; solid lines are from the model. 
  }
  \label{Fig:Chesser}
\end{figure}

Network plasticity may be calibrated based on atomistic simulations to inform the parameters in the model, such as GB dissipation energy and shear coupling. 
Here, atomistic simulations from \cite{chesser2020continuum} of Al STGB bicrystals are used to fit the parameters in NP (\cref{Fig:Chesser}, left). 
The Al bicrystals have periodic boundary conditions in the $y-z$ plan with free surfaces in the $x$ direction and uses the Mishin EAM Al potential \cite{mishin2001structural}. 
The simulation sizes vary per bicrystal, but have cross sections around $3\times3 \:nm^2$ and larger than $16\:nm$ in the interface plane.
The top and bottom slabs along the z direction are both displaced by steps of $0.04 \mbox{\r{A}}$ and relaxed with a conjugate gradient energy minimization \cite{grippo1997globally} at $0\:\mbox{K}$.
There are 500 displacement steps for a total displacement of $4\:nm$ with force tolerance of $10^{-8} \mbox{eV/\r{A}}$ and energy tolerance of $10^{-8}$ in LAMMPS \cite{plimpton1995fast}.

Most of the boundaries exhibit regular shear coupling behavior, with slight fluctuations resulting from atomic level stick-slip behavior.
GB 117 initially has irregular stress drops attributed to GB sliding; though it still produces a desirable sawtooth pattern at higher strains (\cref{Fig:Chesser}, right). 
As the domain size increases to infinity, the stick slip behavior converges to the peak stress with perfect elastic plastic deformation \cite{ivanov2008dynamics}. 
The calculation of $\phi^*_0$ should then only consider the peak stress to mitigate the dependence on domain size and recover $\phi^*_0$ for large grains for use with NP, as it is a mesoscale model.
However, dissipation energy is considered the lower bound for the nucleation energy as disconnection island or loop structure is not considered with its calculation.
However, $\phi^*_0$ is correlated to disconnection mode selection for a uniformly migrating flat GB \cite{chesser2020continuum}.
Comparing the mechanical response of finite atomistic simulations to NP fit with the corresponding parameters, NP is able to capture the average effect of the stick slip behavior. 
In general, atomistic simulations can be run for many different GB geometries to find their respective shear coupling factors, dissipation energies, mobilities, and so on to accurately model GB mediated plasticity in a large microstructures. 

\begin{figure}
    \centering
    \begin{subfigure}[c]{0.24\linewidth}
        \centering
        \includegraphics[height=6cm]{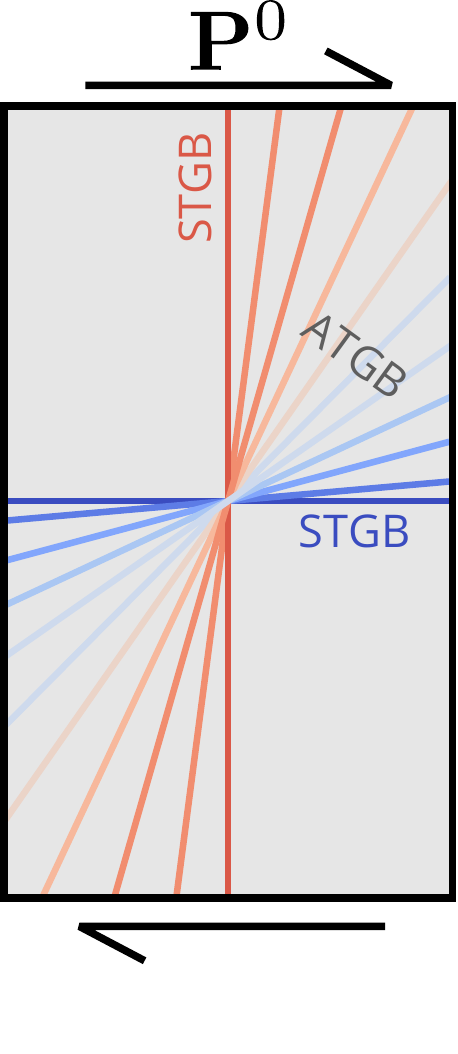}
        \caption{ATGB Schematic}
        \label{Fig:Bicrystal_Rot}
    \end{subfigure}%
    \begin{subfigure}[c]{0.76\linewidth}
        \centering
        \includegraphics[height=6cm]{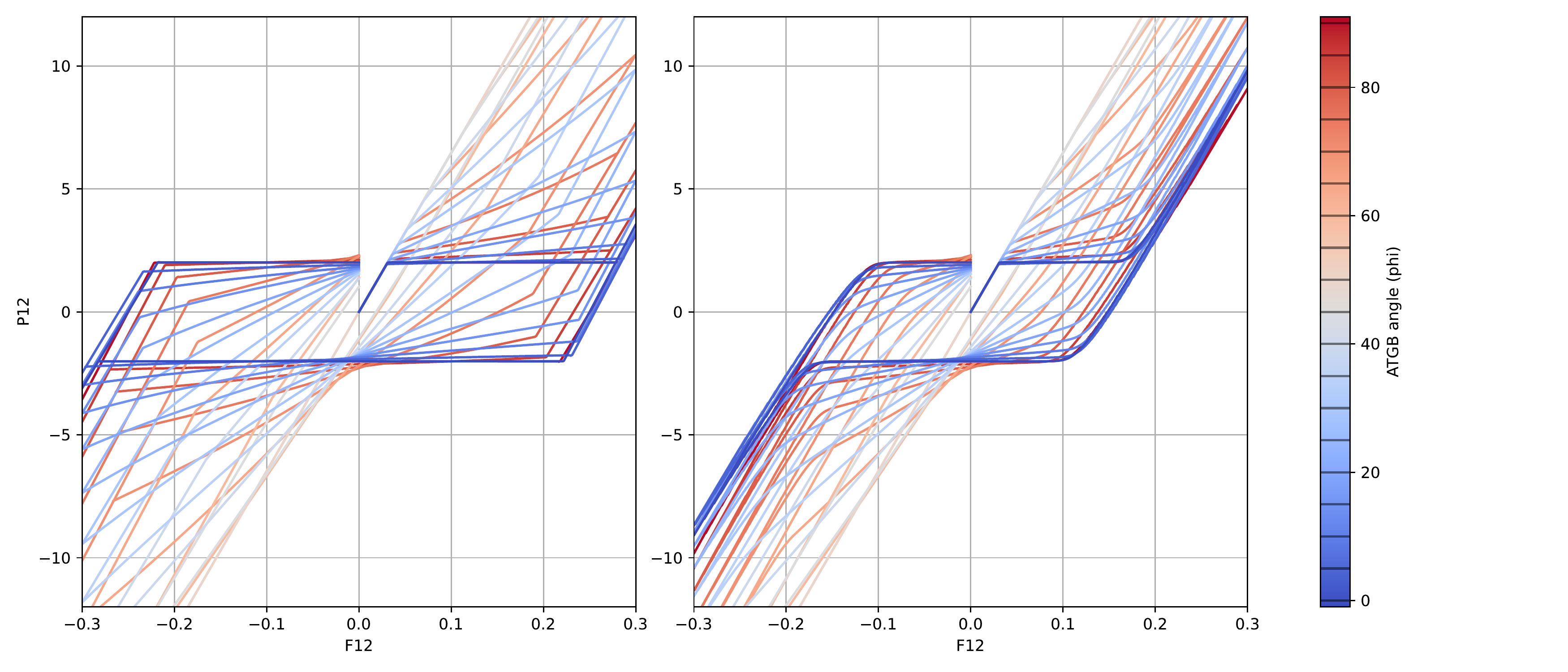}
        \caption{Stress-strain hysteresis without geometric hardening (left) and with hardening (right)}
        \label{Fig:Bicrystal_Rot_hardening_ss}
    \end{subfigure}
    \begin{subfigure}[c]{\linewidth}
        \centering
        \includegraphics[height=3cm]{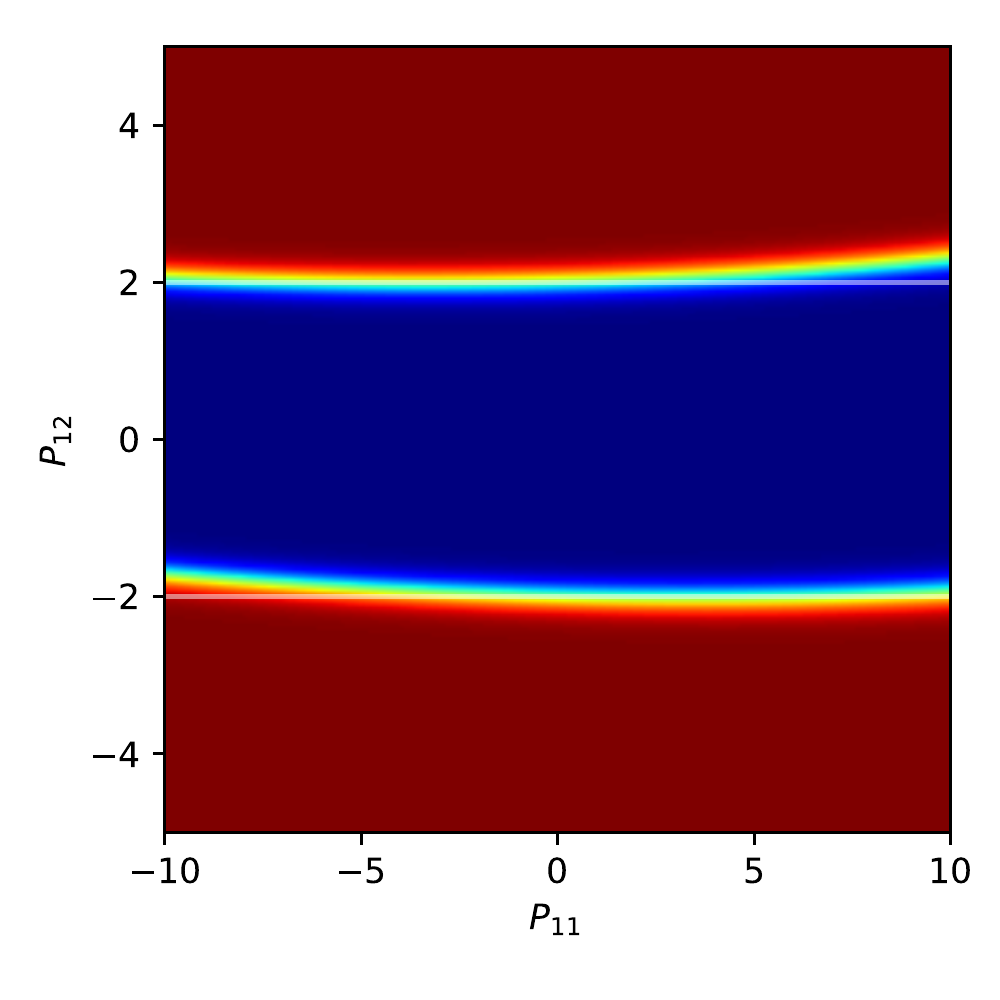}
        \includegraphics[height=3cm]{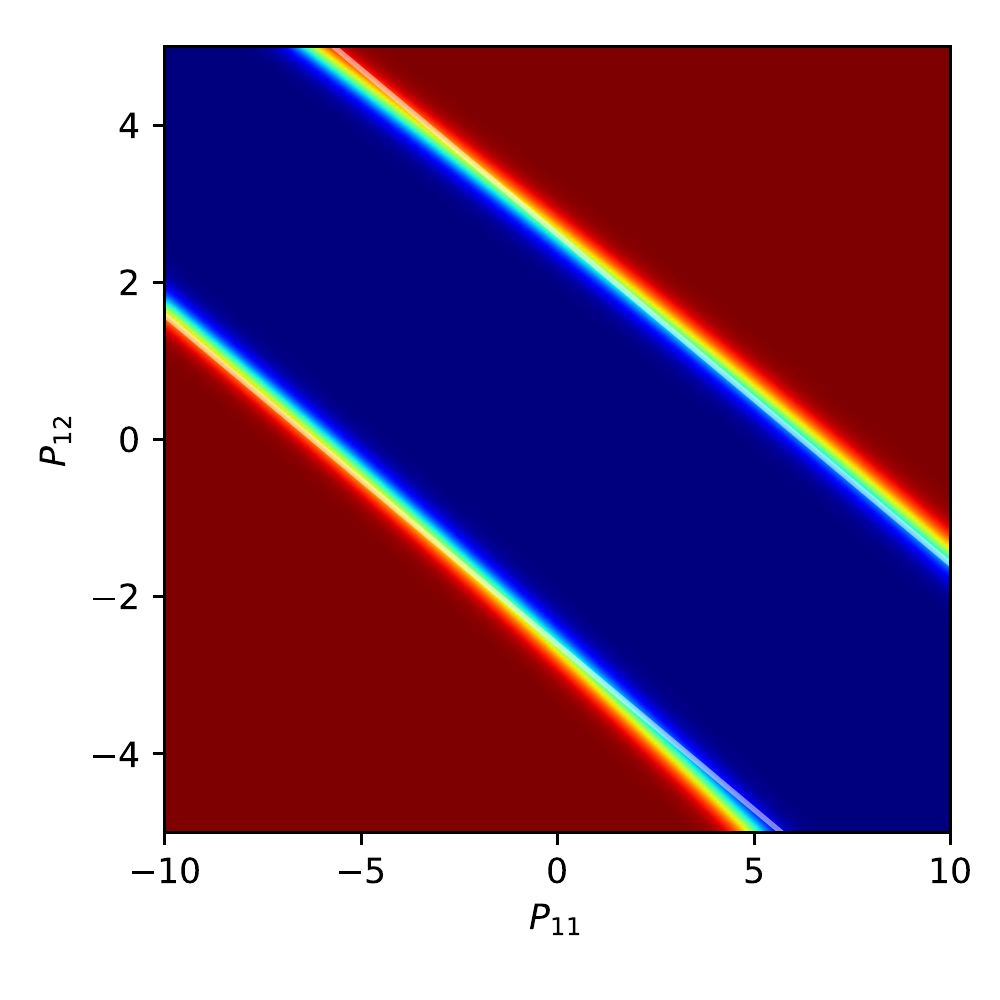}
        \includegraphics[height=3cm]{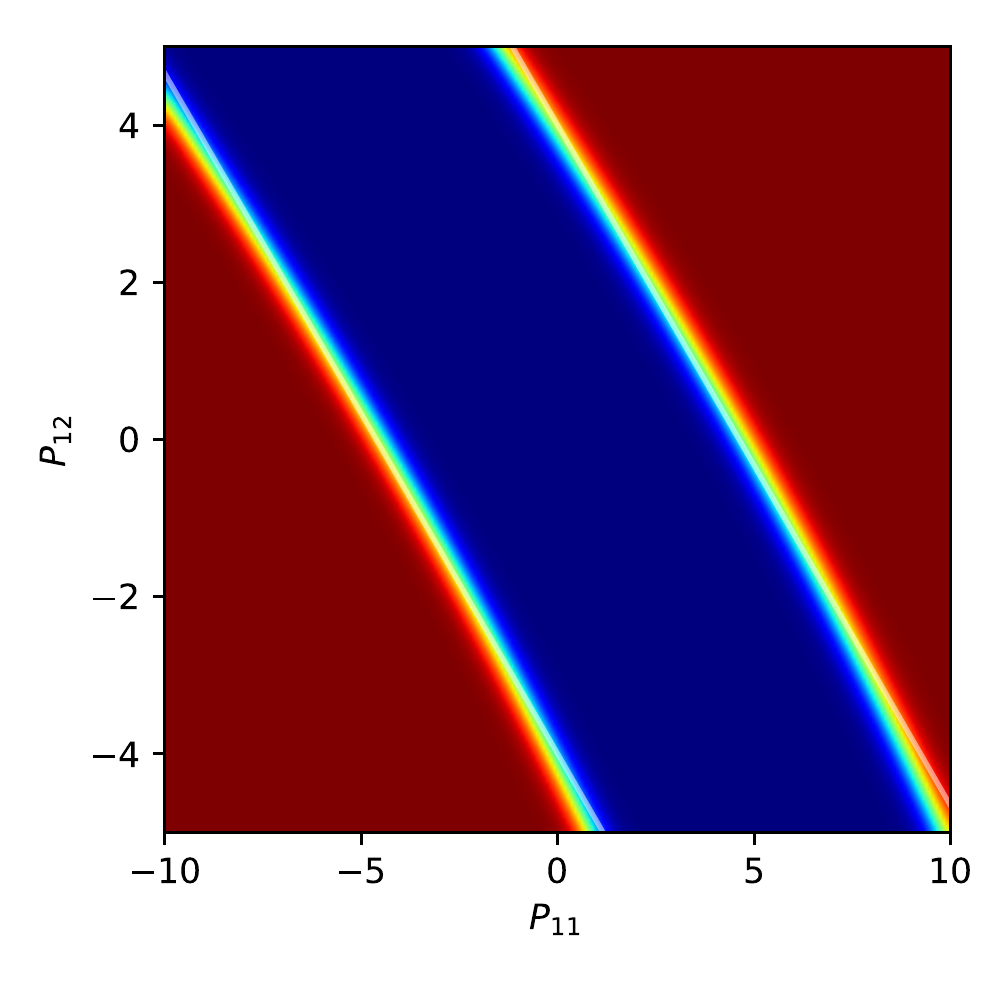}
        \includegraphics[height=3cm]{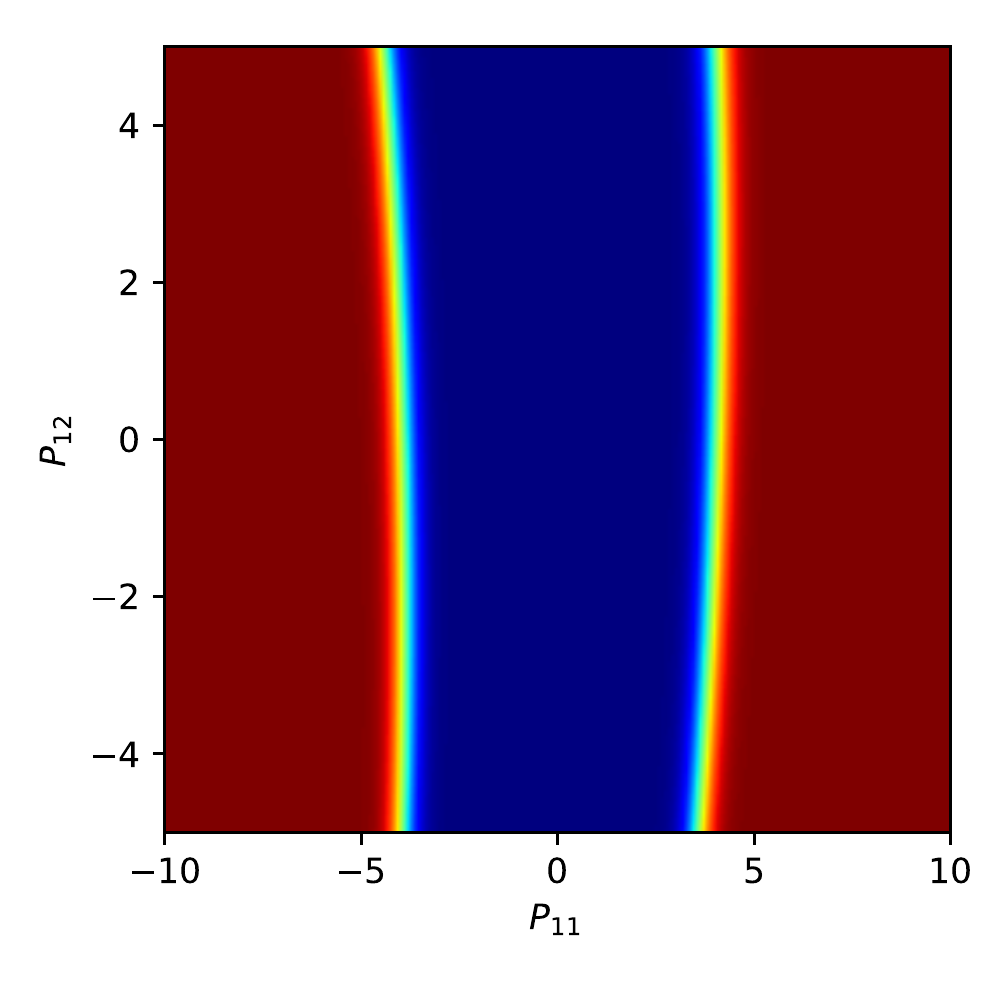}
        \caption{Yield surface for $\phi=0^\circ,20^\circ,30^\circ,45^\circ$}
        \label{Fig:Bicrystal_Yieldsurface} 
    \end{subfigure}
    \caption{
      Stress-strain curves corresponding to boundary motion for various boundary inclinations (a), for elastic-perfectly-plastic deformation (b), and with geometric hardening, (c) yield surfaces at differing inclination angles. The parameters for the hardening potential are: $\phi_1 = 0$, $\phi_2 = 10^{5}$, $m = 11$, $f = 0.5$, and $\mu = 0.1$.
    }
    \label{Fig:Bicrystal_example}
\end{figure}

\subsection{Asymmetric boundaries and hardening}

Simple Cu bicrystals with varying STGB and ATGB boundaries (\cref{Fig:Bicrystal_Rot}) are explored to show their effects with and without the hardening potential. 
Shear coupling for ATGBs is approximated by rotation of the STGB shear coupling matrix as a function of the GB inclination angle as seen in \cref{ATGBshearcoupling}. 
When the GB inclination angle increases, the yield of the material increases accordingly along with material hardening (\cref{Fig:Bicrystal_Rot_hardening_ss}). 
Yield increase of ATGBs is well-known and likely results from strong interactions that block dislocations \cite{read1950dislocation,trautt2012coupled}. 
The rotated shear coupling matrix also reproduces elastic anisotropy which requires extra stresses to move a boundary, and effectively hardens the material \cite{trautt2012coupled}.
Shear coupling in a bicrystal produces a columnar yield surface, while elastic anisotropy between grains $p$ and $q$ gives small deviations from the linear driving force of shear coupling (\cref{Fig:Bicrystal_Yieldsurface}). 
Furthermore, changes in the GB inclination angle cause a rotation of the yield surface, such that a boundary with $\phi = 45 ^{\circ}$ does not move under pure shear, and when $\phi = 0 ^{\circ}$ does not move with pure tension or compression.  

GB motion and its shear coupling factor directly corresponds to the mechanical response from GB motion, while the boundary's dissipation energy changes yield. 
If an STGB boundary moves freely without interactions at triple points, the idealized migration produces perfect elastic plastic deformation until a grain is consumed. 
When a grain's volume is consumed, the boundary stops moving and plastic deformation stops, and subsequent deformation is accommodated by elasticity. 
If hardening is considered, the consequence is reduced ductility caused by restricting GB motion. 
When geometric hardening is introduced, this mimics the effect of pinning at triple points, and as GB motion slows to a stop, plastic deformation ceases.
Dislocations that account for incompatibility caused by GB motion creates climb hardening supplemental to hardening of ATGBs (recall \cref{Fig:hardening}). 

This example demonstrates the effect of ATGB shear coupling: namely, the increase in yield strength and ``climb'' hardening induced by incompatible motion and build-up of back stress.
It should be emphasized that the ATGB effect is not the result of a hardening model, but is captured naturally even in the elastic-perfectly-plastic case.
This example also shows the behavior of geometric hardening, which implicitly demonstrates the effect of triple point pinning.
Additional mechanical behavior can be readily captured through the inclusion of more sophisticated hardening potentials, which prompts the need for additional atomistic calibration data.

\subsection{Twinning and triple point plasticity} \label{sec:example_subsets}

\begin{figure}
  \begin{subtable}{0.5\linewidth}
    \centering\footnotesize
    \begin{tabularx}{\linewidth}{lX}
      \toprule
      \multicolumn{2}{c}{Elastic Constants (Copper \cite{berbenni2013micromechanics})}\\
      \midrule
      $C_{11} = 169.3$ GPa & $C_{12} = 122.5$ GPa \\
      $C_{44} = 76$ GPa  \\
      \midrule
      \multicolumn{2}{c}{Geometric hardening}\\
      \midrule
      $\phi^*_{h_{2}} = 2\times10^{-5}$ &
      $m = 11$ \\
      $f = $ 0.005 / 0.00005 &
      $\mu = 13.33$  \\
      \midrule
      \multicolumn{2}{c}{Shear coupling (FCC)}\\
      \midrule
      \begin{minipage}{0.5\linewidth}
      \begin{equation*}
          \mu =
        \begin{cases}
          1.0 & 0\geq\theta_r>36.9^{\circ}  \\
          0.75 &  36.9^{\circ}\geq\theta_r>50.48^{\circ}\\
          0.5  & 50.48^{\circ}\geq\theta_r
        \end{cases}
      \end{equation*}
      \end{minipage}
      &
      Rot. sym. n=2 \\
      \bottomrule
    \end{tabularx}
    \vspace{0.4cm}
    \caption{Table of parameters}
    \label{Table:params}
  \end{subtable}%
  \begin{subfigure}{0.5\linewidth}
    \centering
    \includegraphics[width=\textwidth]{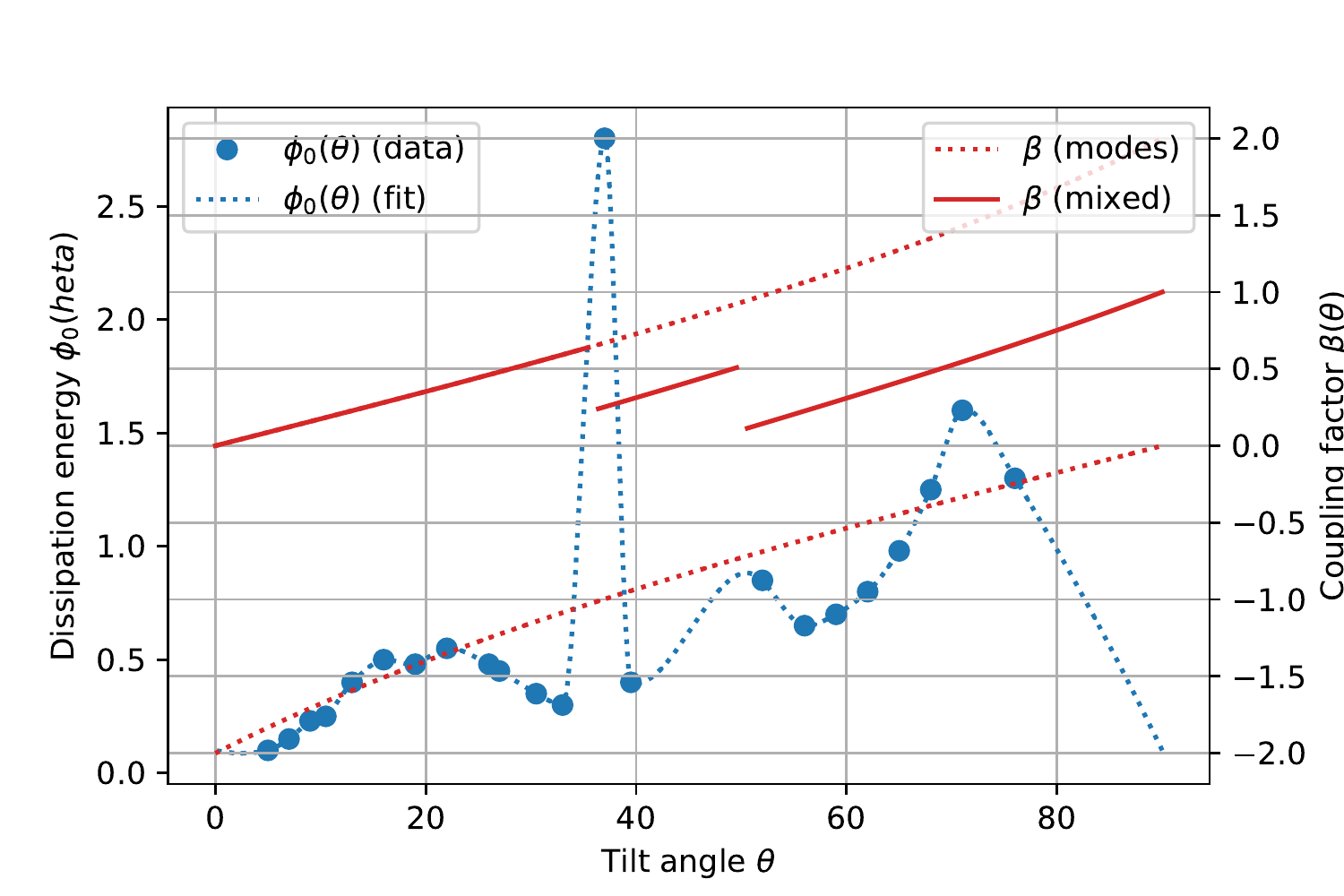}
    \caption{Dissipation energy and shear coupling factor}
    \label{Fig:DissipationDist}
  \end{subfigure}
  \caption{Approximated values for unknown material parameters for NP applied to EBSD microstructure}
\end{figure}

To determine the effect of NP in microstructures that are more complex than bicrystals, yet with a tractable level of complexity, two representative microstructure subsets are considered taken from experimental observation.
Copper EBSD Data from Randle \etal\ \cite{RANDLE20082363} is reconstructed into the NP framework to predict the mechanical properties of the microstructure. 
This reconstruction is a simplification of geometric aspects of the grains, as complex boundaries are replaced by effectively planar boundaries.

NP requires an extensive number of parameters. 
Standard elastic constants are used, and orientation information from EBSD is used to rotate the elastic modulus tensors accordingly.
In order to predictively model a microstructure, NP parameters must be informed either through atomistic techniques or calibration to macroscale response.
In absence of this necessary work, rough estimates of the necessary parameters are used.
The shear coupling factor is estimated by finding the Rodrigues vector between two neighboring grains and crystallographic orientation given in the Bunge convention. 
The disorientation angle between two grains, $\theta$ is measured and related to $\beta$ through the geometric formula,
\begin{equation}
  \label{beta}
  \beta =  2\mu\tan\left(\frac{\theta}{2}\right)-2(1-\mu)\tan\left(\frac{\pi}{n}-\frac{\theta}{2}\right),
\end{equation}
where $\mu$ is a mode-mixing factor and $n$ is the order of rotational symmetry (values tabulated in table \ref{Table:params}). 
The form of \cref{beta} was originally suggested in \cite{molodov2018grain} as a general estimation for finding the shear coupling factor for different STGBs in materials by varying fitting parameters $\mu$ and $n$. 
Shear coupling factors can drastically change based on the effects of triple junctions, ATGBs \cite{trautt2012coupled}, and loading mode \cite{aramfard2014influences}, which \cref{beta} does not consider.
ATGBs angles are estimated by the angle between the vector between two triple points and the Rodrigues vector, and are then used to rotate the STGB shear coupling tensor accordingly.

The dissipation potential is also dependent on the crystallography between grains. 
Relatively little work is available from which to draw values for the 
dissipation energy, as most atomistic work has been focused on finding the mobility.
For a rough estimation of $\phi_0^*$, Al dissipation energy for $[100]$ STGBs from data in \cite{chesser2020continuum} is used (\cref{Fig:DissipationDist}).
In general, the relationship between dissipation energy and boundary character is more complex; incorporation of full five degree-of-freedom information into dissipation energy estimation will be left to future work.
It has been observed that GB mobility is correlated to the dissipation energy associated with the GB at low temperatures \cite{chesser2020continuum,yu2019survey}, such that large dissipation energies have low mobility and vice versa. 
Therefore, to estimate mobility, $\phi_1^* = \eta\phi_0^*$ is used where $\eta$ is a fitting parameter. 
The mobility of Cu GBs is generally on the order of $10^{3}\:\mbox{M}/\mbox{GPa}/\mbox{s}$ \cite{chesser2018understanding}, and so $\eta = 0.0005$ is used to approximate this order of magnitude.

\begin{figure}
  \centering
  \begin{subfigure}{.25\textwidth}\centering
    \includegraphics[height=4.5cm]{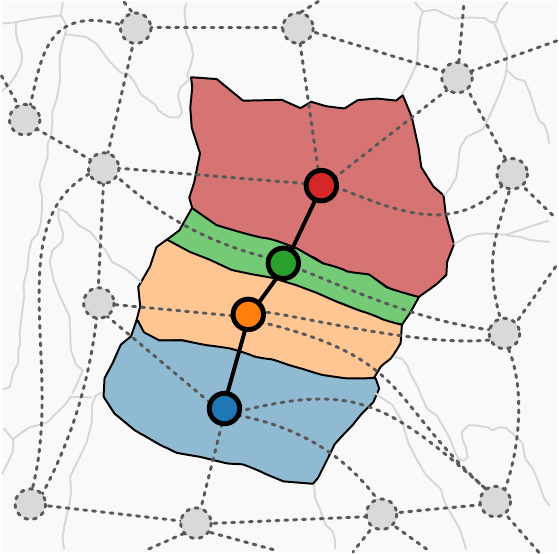}
    \caption{Twin subset}
    \label{fig:EBSD_twin}
  \end{subfigure}\hfill%
  \begin{subfigure}[c]{.25\textwidth}\centering
    \includegraphics[height=4.5cm]{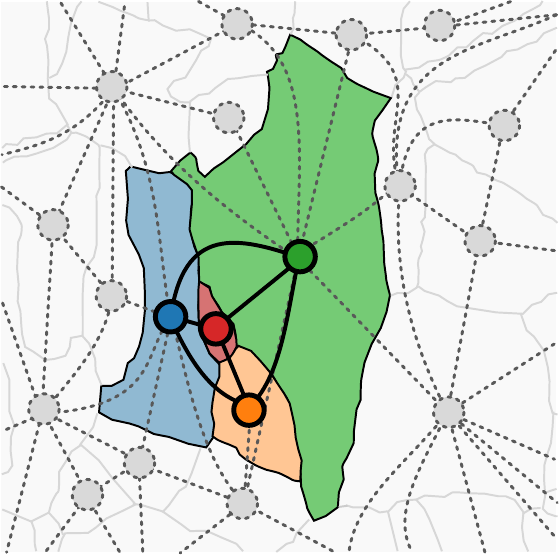}
    \caption{Random subset}
    \label{fig:EBSD_random}
  \end{subfigure}\hfill%
  \begin{subfigure}[c]{.42\textwidth}\centering
      \includegraphics[height=4.5cm,clip,trim=0cm 0.5cm 0cm 0cm]{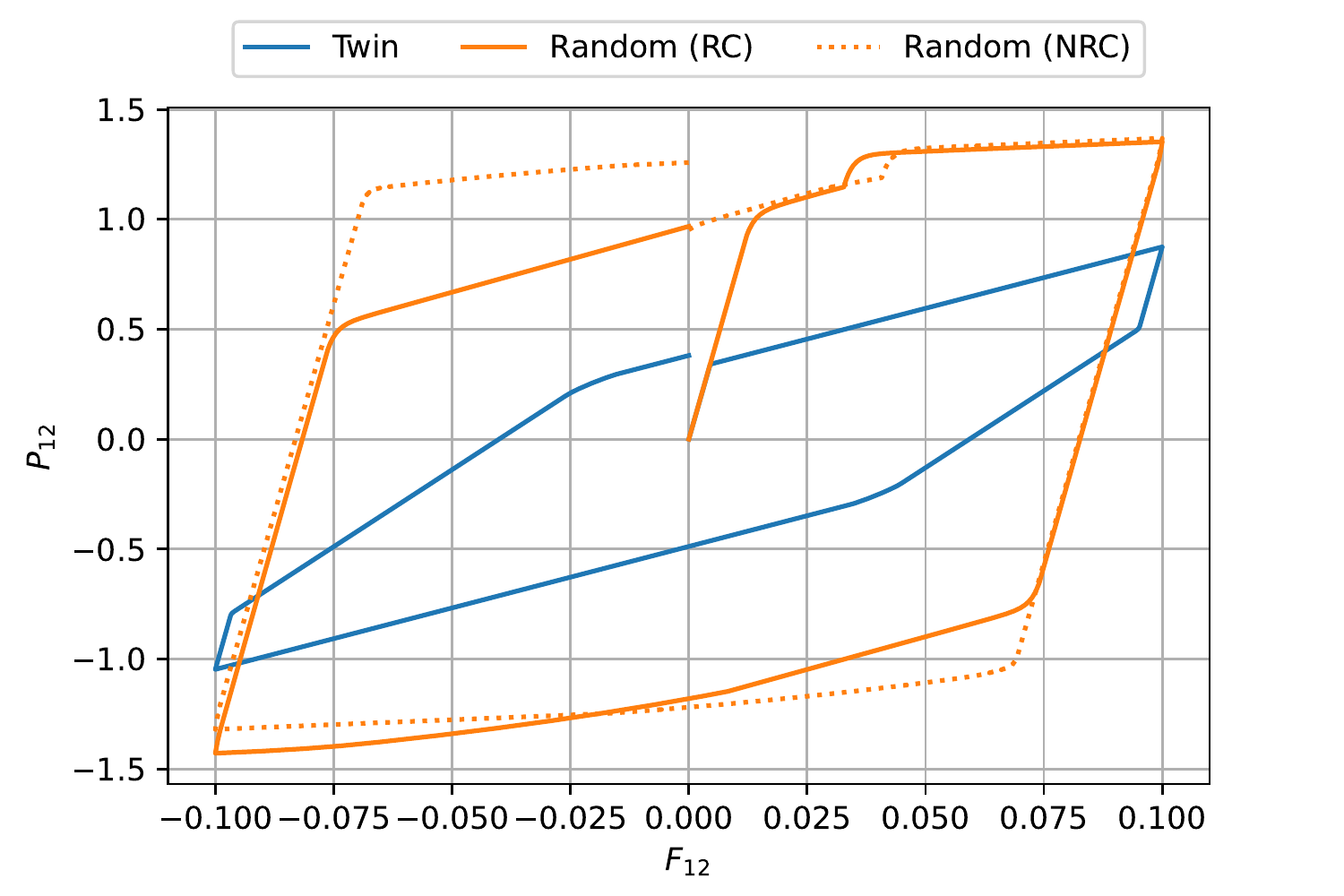}
      \caption{Stress-strain response}
    \label{fig:stressstrain_twin_random}
  \end{subfigure}
  \begin{subfigure}[c]{.45\textwidth}\centering
    \includegraphics[height=4.5cm]{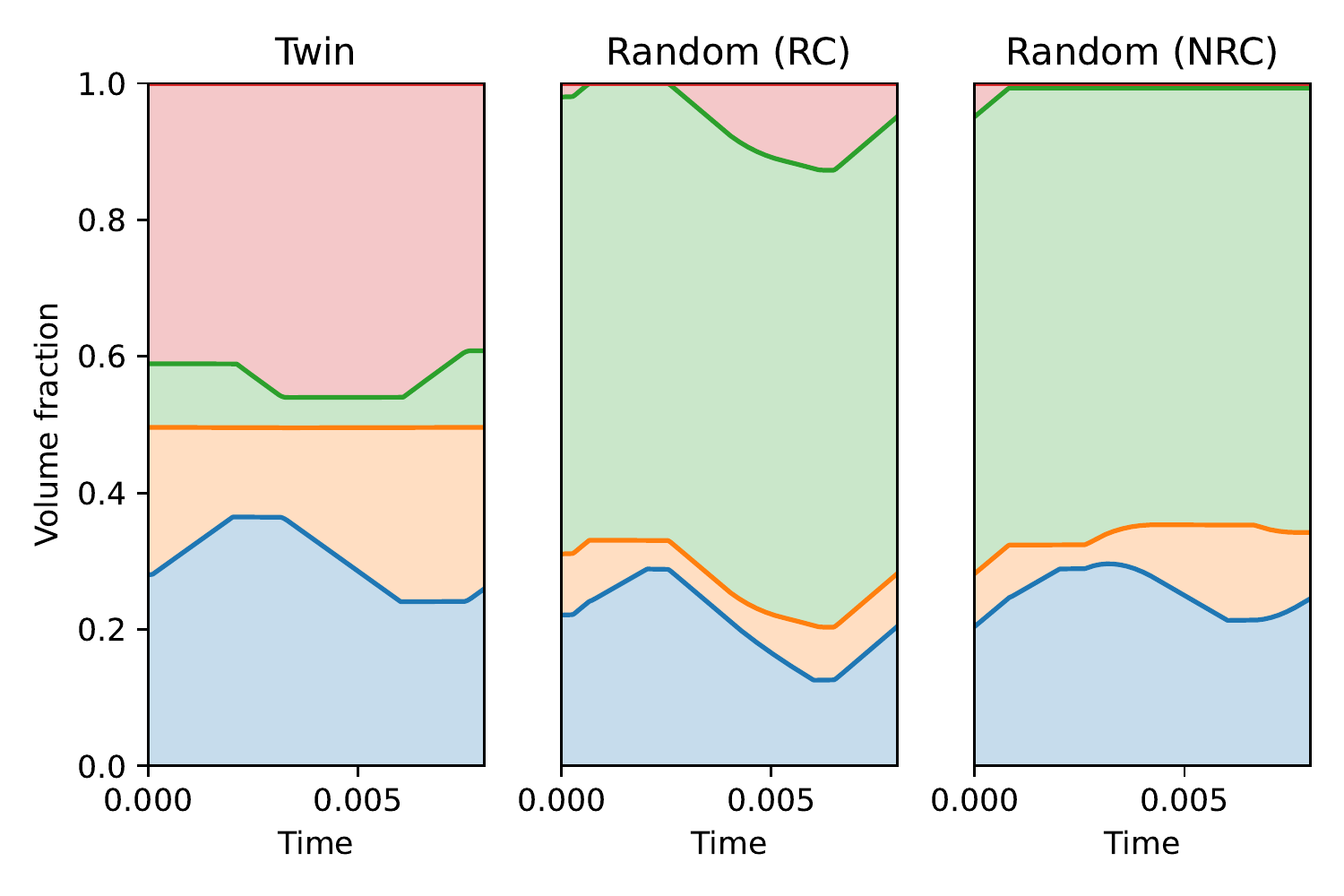}
    \caption{Volume fractions of twin (left), RC-random (center), and NRC-random (right)}
    \label{fig:volfrac_twin_random}
  \end{subfigure}%
  \begin{subfigure}[c]{.25\textwidth}\centering
    \includegraphics[height=4.5cm]{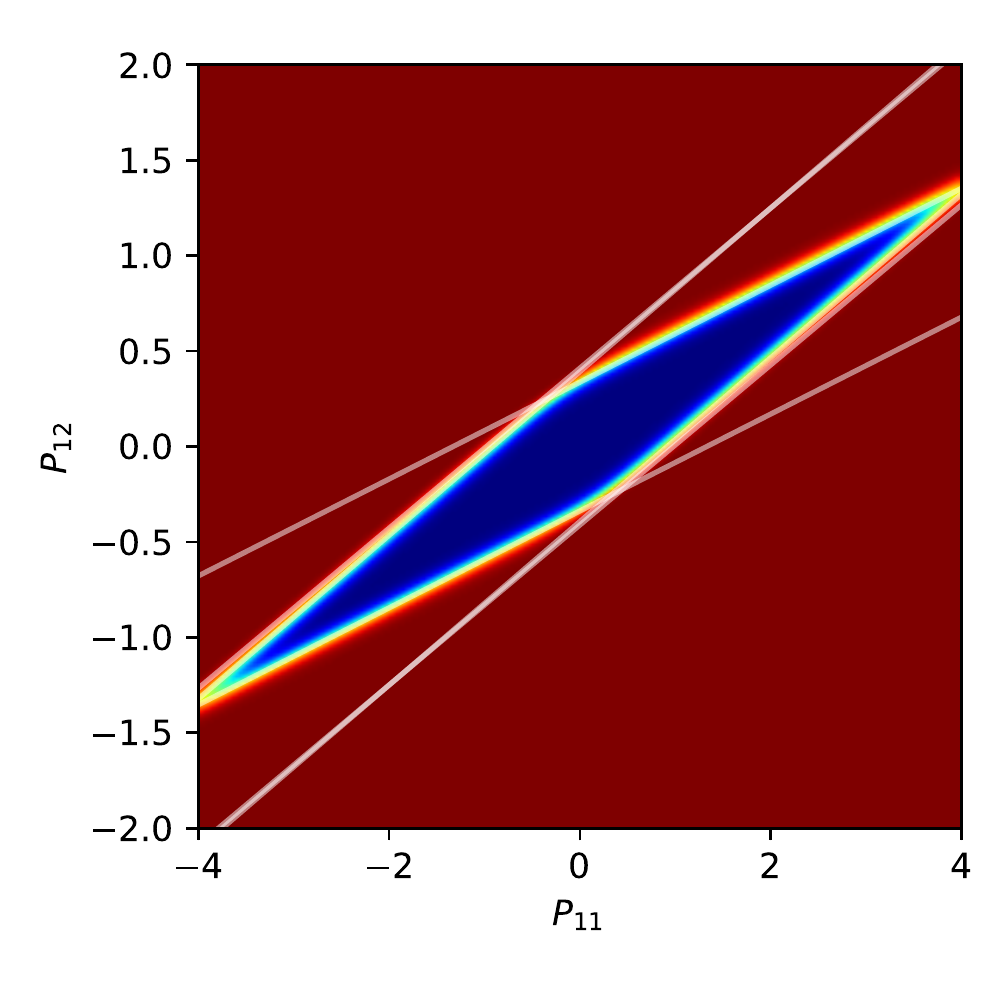}
    \caption{Twin yield surface }
    \label{fig:ys_twin}
  \end{subfigure}%
  \begin{subfigure}[c]{.25\textwidth}\centering
    \includegraphics[height=4.5cm]{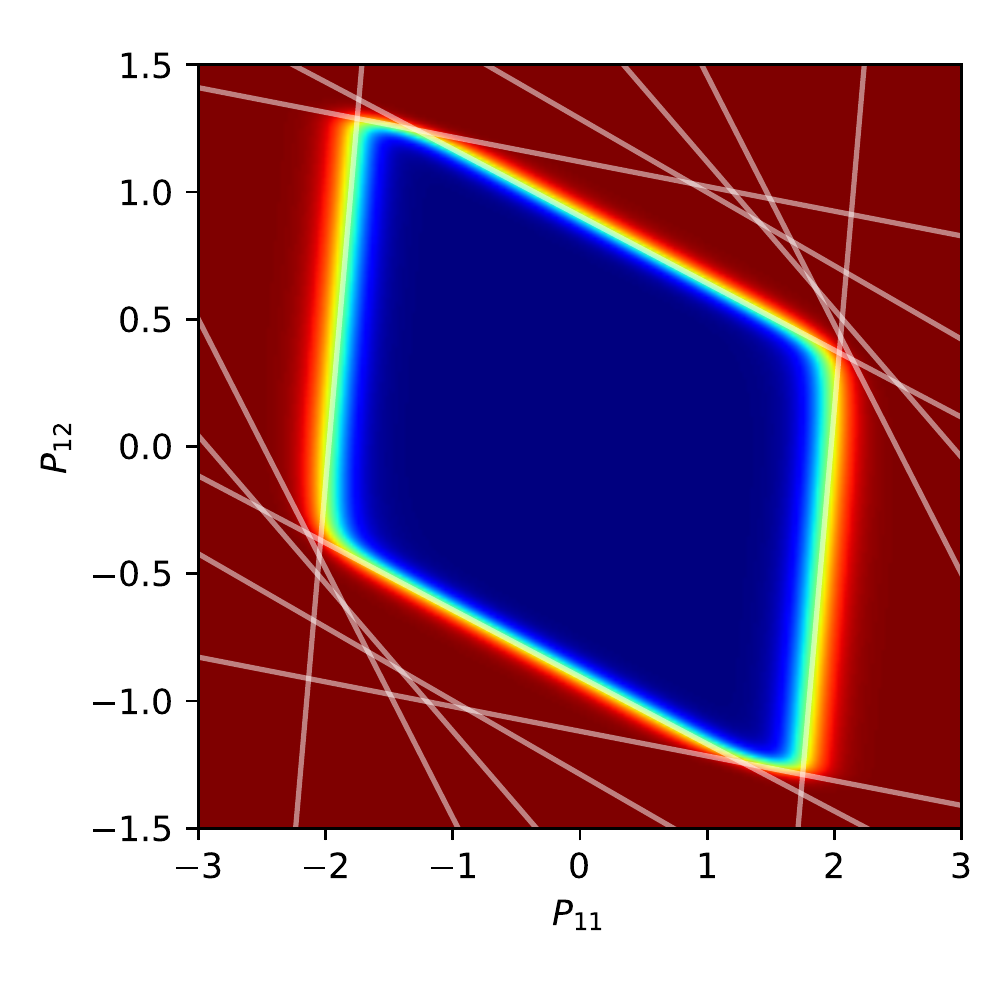}
    \caption{Random subset yield surface}
    \label{fig:ys_random}
  \end{subfigure}
  \caption{
    Demonstration of the effect of grain boundary character on network-plastic behavior by activating microstructure subsets.
    (a) approximately twin boundaries with low dissipation energy and high shear coupling factor, and (b) a random collection of asymmetric tilt grain boundaries.
    (c) Hysteresis loop is measured for the random and twin subsets
    (d) volume fractions during the loading cycle
    (e-f) yield surfaces
  }
\end{figure}

Two subsets of the EBSD data are considered: a laminate (``twin'', \cref{fig:EBSD_twin}) comprising of twin boundaries, and a subset of randomly oriented boundaries (``random'', \cref{fig:EBSD_random}). 
The twin boundaries in the laminate have low dissipation energies and large shear coupling factors characteristic of $\Sigma 3$ boundaries recovered by the present methodology to estimate these values. 
In FCC materials $\Sigma 3$ boundaries are often associated with twin boundaries and indirectly responsible for desirable material property changes \cite{randle2004twinning}. 
This behavior is typically found in low stacking fault energy materials, such as Cu \cite{lin1995influence}. 
As the proportion of twined boundaries increase, the result are new triple junctions due to restrictions imposed by twinning and cause an overall harder more ductile material \cite{randle1999mechanism}.
Generally, increasing the fraction of Low $\Sigma$ or low angle boundaries produces more desirable material properties in conjunction with reducing networks of random boundaries \cite{tsurekawa1994grain,randle2004twinning}.
Twinned boundaries' shear coupling behavior and STGBs have been researched in depth and are reasonably characterized by \cref{beta}.  

The plastic hysteresis curves for both subsets are measured by subjecting to simple shear with magnitude following one cycle of a triangle wave (\cref{fig:stressstrain_twin_random}).
While rate dependency is included in this model, the strain rate is small enough so that there are no strain rate effects.
In the twin subset, plastic behavior is similar to that measured for an ATGB bicrystal. 
Hardening of the ``climb'' type (i.e. induced by incompatibility of the ATGB shear coupling) is observed due to the misorientation of the twin boundaries from the loading direction.
There is some degree of tension/compression asymmetry, which is confirmed in some measurements of twin migration \cite{tucker2015quantifying}; however, the magnitude of initial yield is not similarly affected.
The ductility of the twin boundary is mediated through the motion of the boundaries in the graph (\cref{fig:volfrac_twin_random}, left).
Unsurprisingly, motion is dominated by the top (red-green) and bottom (yellow-blue) boundaries, while the middle boundary (yellow-green) does not move.
Under positive shear, the blue and green grains grow, while under negative shear, the yellow and red grains grow.
In the limit, this would correspond to de-twinning of the sample.

The random subset case demonstrates a higher yield than the twin subset case (as expected), although the post-yield behavior is, interestingly, more ductile.
In the random subset case, initial plastic deformation is accommodated through the shrinkage of the red grain into the blue grain (\cref{fig:volfrac_twin_random}, right). 
This effects a coarsening of the microstructure, as the red vertex effectively disappears from the graph and the results is a triple point between the blue, green, and yellow grains.
The effect is a step in the stress-strain curve, with an initial yield and then a brief return to elastic flow once the red volume is depleted.
There are two modeling options at this point: (1) forbid the red grain from re-appearing by imposing a lock on any volumes that have decreased to zero, or (2) leave the red grain in the network and allow it to regrow when favorable.
Both options are considered here, and are referred to as the ``re-creation'' (RC) and ``no-re-creation'' (NRC) cases, respectively.
After the initial disappearance of the red grain, nearly perfect shear coupling is observed in the blue-yellow boundary, causing the continued growth of the blue grain.
  Up until this point, the RC and NRC cases behave identically, as expected.
  When the loading is reversed, the RC case permits the red grain to grow.
  On the other hand, the NRC case proceeds as if the red grain no longer exists.
  As a result, the yield stress for the NRC case is considerably higher than that for the RC case.
  This demonstrates the hardening effect of coarsening on GB-mediated plasticity.
It is interesting to note that the orange-blue boundary has significant ductile response even compared to high shear coupling factor of the twin boundaries.
This is due to the ATGB approximation where the $\Fgb$ is nearly $90^{\circ}$ and closely acts like a STGB under an applied shear along with a large shear coupling factor.
However, the blue-orange boundary also has a large dissipation energy and is not likely to activate compared to twin boundaries.

The yield surfaces corresponding to initial yield are constructed for both subsets (\cref{fig:ys_twin,fig:ys_random}).
Like \cref{Fig:Bicrystal_Yieldsurface}, only a 2d slice is plotted in $\mathrm{P}_{11}-\mathrm{P}_{12}$ space, with $\mathrm{P}_{22}=0$, and the boundary is smoothed slightly to prevent aliasing.
The white lines correspond to yield surfaces for individual boundaries, two for each boundary, and are semitransparent so that overlapping lines are visible.
In the twin case, the blue-yellow and red-green boundaries are the same, and together with the yellow-green boundary, produce a highly elongated yield surface.
This corresponds to the intuition that twins yield easily under certain loadings and not others.
On the other hand, the yield surface for the random subset case has a lower aspect ratio, but an overall greater magnitude in most directions.
The difference between the yield surfaces highlights the effect of microstructure on plastic behavior.

\subsection{Grain boundary engineered microstructures}

The final example in this work is the application of NP to large-scale microstructures containing $\mathcal{O}(500)$s of grains with 3-4 times as many boundaries.
To highlight the effect of GB-mediated plasticity, samples are considered from GB engineered microstructures as well as as-processed microstructures.
GB engineering is the process of controlling the crystallographic, chemical, or spatial distribution of GBs \cite{watanabe1999control} to result in materials tailored for a specific purpose.
The most well-known and basic example of GB engineering is the Hall-Petch effect \cite{cordero2016six}, which is the inverse relationship between yield strength and grain size.
NP cannot directly recover the Hall-Petch effect because it is caused by interactions between dislocations and GBs, which is not modeled by NP. 
Aside from grain size, the GB character distribution has also been shown to greatly affect material strength \cite{watanabe1999control}. 
However, NP exhibits a softening effect when more grains are included, which is reminiscent of the inverse Hall-Petch effect \cite{naik2020hall}, the decrease in yield stress corresponding to increased boundary content.
This behavior may be interpreted as a kind of size effect, but it should be noted that it is controlled by the total number of grains and boundaries, which does not necessarily equivocate to size.
\begin{figure}
  \centering
  \includegraphics[width=0.8\linewidth]{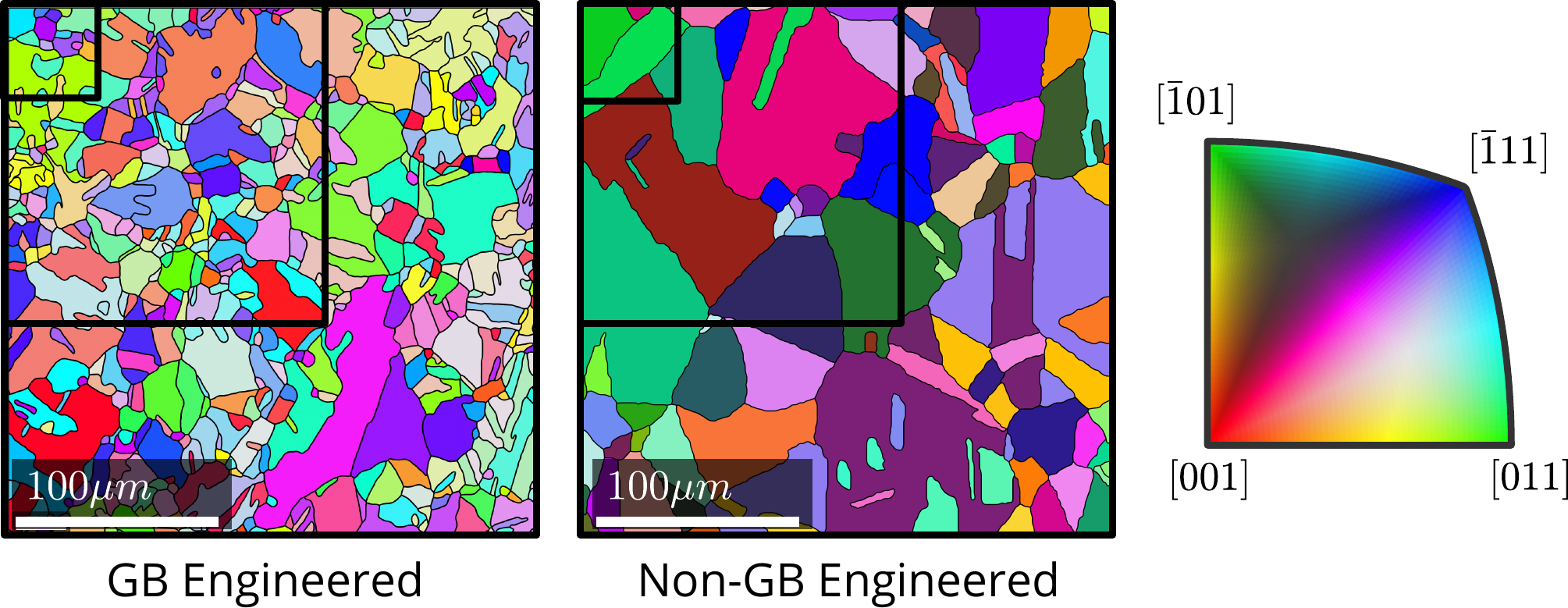}
  \caption{Engineered and non-engineered boundaries with small, medium, and large subsets.}
  \label{fig:microstructures}
\end{figure}

The EBSD samples modeled are oxygen free high conductivity copper for both GBE and non-GBE from data sets. 
Analysis of the EBSD data done by Randle \etal\ \cite{RANDLE20082363,randle2006mechanisms} showed that the main differences in GBE sample over non-GBE sample was low energy $[110]$ ATGBs and STGBs cause a larger portion of $\Sigma3^n$ triple junctions. 
The GBE sample had a large portion of $\Sigma 3$ boundaries created through two $\Sigma3$ boundaries meeting a triple points generating $\Sigma9$ boundaries. 
When $\Sigma9$ boundaries meet another $\Sigma3$ it is more likely to make a $\Sigma3$ than a $\Sigma27$ boundary. 
Many of the ATGBs are vicinal to low index planes, which increase intergranular corrosion resistant boundaries. 
These interactions dominate the resulting GBE properties as opposed to $\Sigma 3$ boundaries in the $[111]$ plane. 
On the other hand, twinning in both of the GB engineered and non-engineered data typically occurs along the $[111]$ plane.
This is common for FCC materials, since the $[111]$ plane represents the densest planes in FCC lattices. 
It was shown in \cite{rohrer2006changes} that this method of GBE has a high likelihood to increase ductility, via experiments with brass, though the mechanical response of the EBSD data is not known. 
Most of the boundaries are faceted, and result in only a proportion of the boundary being $\Sigma3$; though intergranular dislocation transmission will still be restricted. 
The process of GB engineering in which $\Sigma 3$ proportion is increased is indicative of stress induced GB migration or recrystallization during high temperature annealing \cite{randle1999mechanism}, though the method of GB engineering mechanism is unknown. 

\begin{figure}
  \begin{subfigure}{0.5\linewidth}
    \includegraphics[width=\linewidth]{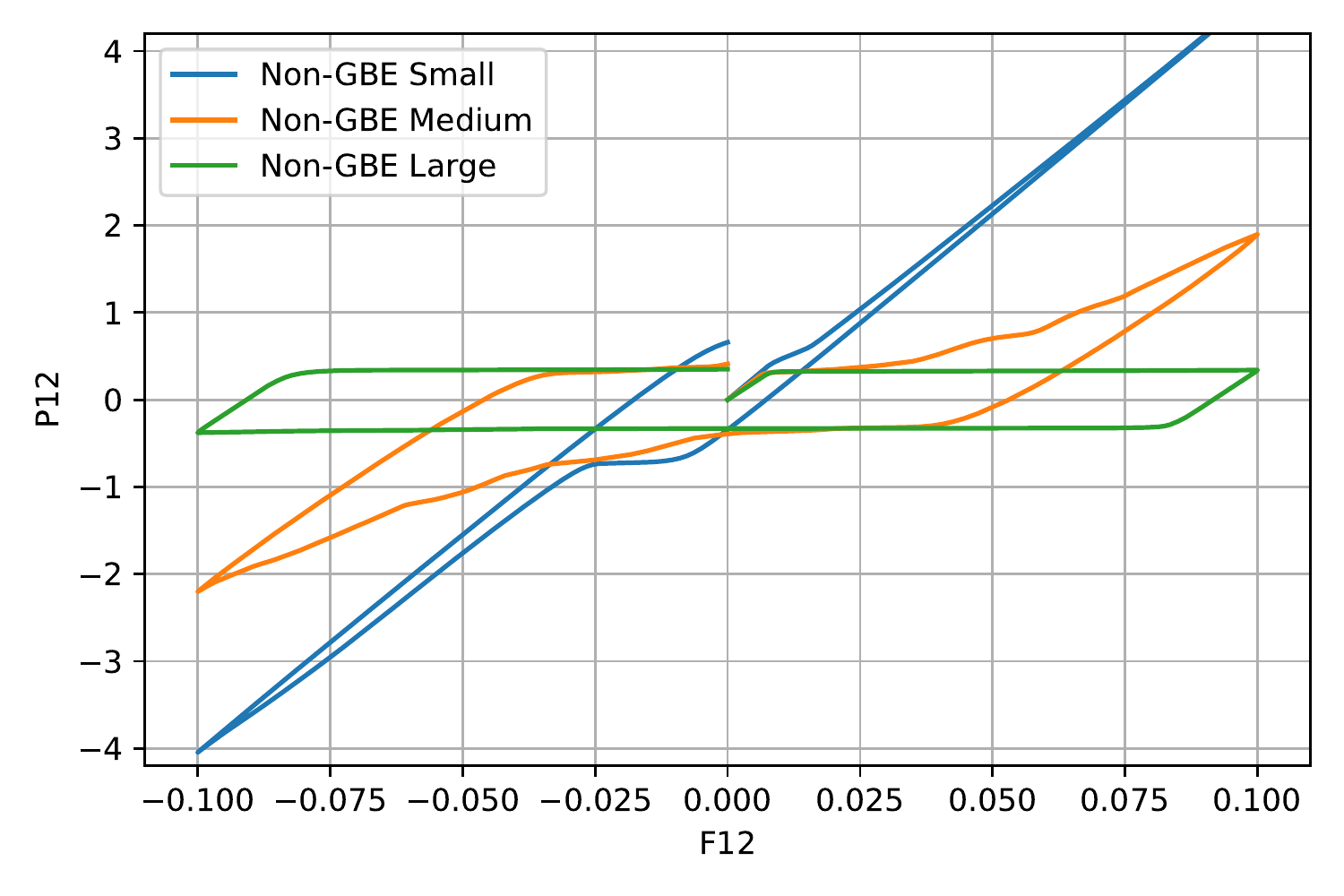}
    \caption{Non-GBE microstructure}
    \label{fig:GrainData_GB_NONeng_SS}
  \end{subfigure}%
  \begin{subfigure}{0.5\linewidth}
    \includegraphics[width=\linewidth]{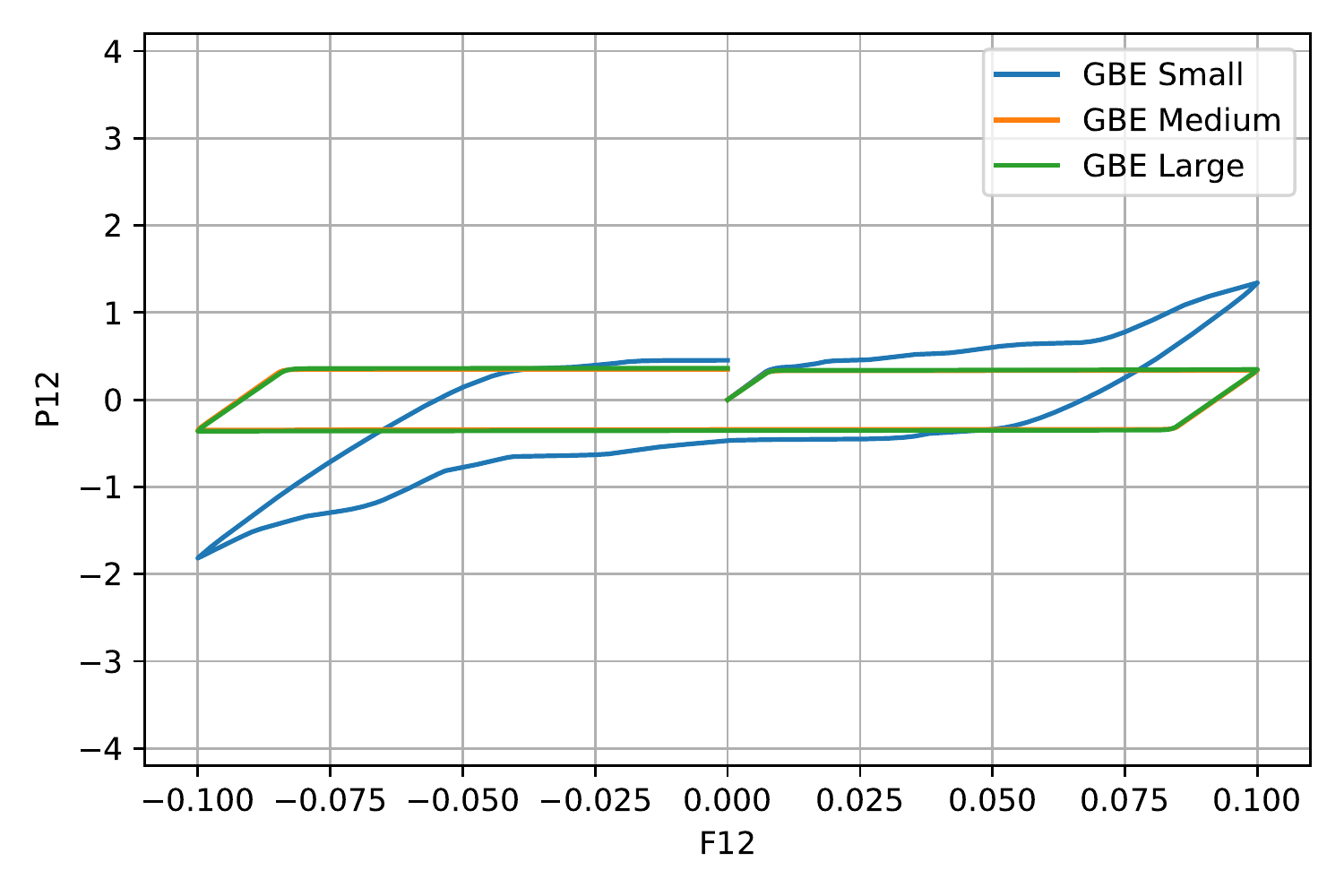}
    \caption{GBE microstructure}
    \label{fig:GrainData_GBeng_SS}
  \end{subfigure}
  \caption{Stress-strain response to pure shear loading for small / medium / large subsets of GBE and Non-GBE microstructures. }
\end{figure}

The NP parameters were estimated using the same process described in \cref{sec:example_subsets}.
However, the hardening parameters are given by values: $\phi^*_{h1} = 0$, $\phi^*_{h2} = 2\times10^{5}$, $m = 11$, $\mu = 0.1$, and $f \sim\frac{1}{|\mathbb{V}(\graph)|}$.
The non-GBE and GBE microstructure are broken up into three samples that have differing size and number of grains, which range from 7-107 grains and 24-565 grains for the non-GBE and GBE data sets respectively (\cref{fig:microstructures}). 
Each sample has the same dimensions of $48\mu m \times48\mu m$, $156\mu m \times156\mu m$, and $260\mu m \times 260\mu m$ described as small, medium, and large sizes respectively.
These samples are examined to find the effect of the number of grains on the mechanical response under an elastic hysteresis (\cref{fig:GrainData_GB_NONeng_SS,fig:GrainData_GBeng_SS}).

The small non-GBE sample exhibits almost no plastic behavior, compared to the small GBE sample which exhibits much greater ductility. 
This is consistent with the known effect of GBE on mechanical response; however, in NP, it is more likely that this is due to the differing numbers of grains in the samples.
The response of the medium non-GBE sample is similar to that of the small GBE sample (again, most likely due to comparable numbers of grains); however, there are still noticeable differences in the hardening behavior due to the differing characters of the GBs.
This is likely due to the differences in $\Sigma3$ boundaries: the small data set for non-GBE contains $8.9\%\: \Sigma3$ boundaries compared to $35.6\% \:\Sigma3$ boundaries in the GBE sample. 
The large non-GBE, medium GBE, and large GBE stress-strain curves have converged to an effectively elastic-perfectly-plastic hysteresis with no noticeable hardening.
In general, it is not likely that this convergent behavior is realistic.
This is because the present NP accounts for shear coupling of boundaries only, and not for dislocation-mediated plasticity.
The similarities may result from limited modeling of the shear coupling factor, GB mobility, and dissipation potential which vary greatly on GB character, but not being modeled.

\begin{figure}
  \begin{subfigure}{0.5\linewidth}
    \includegraphics[width=\linewidth]{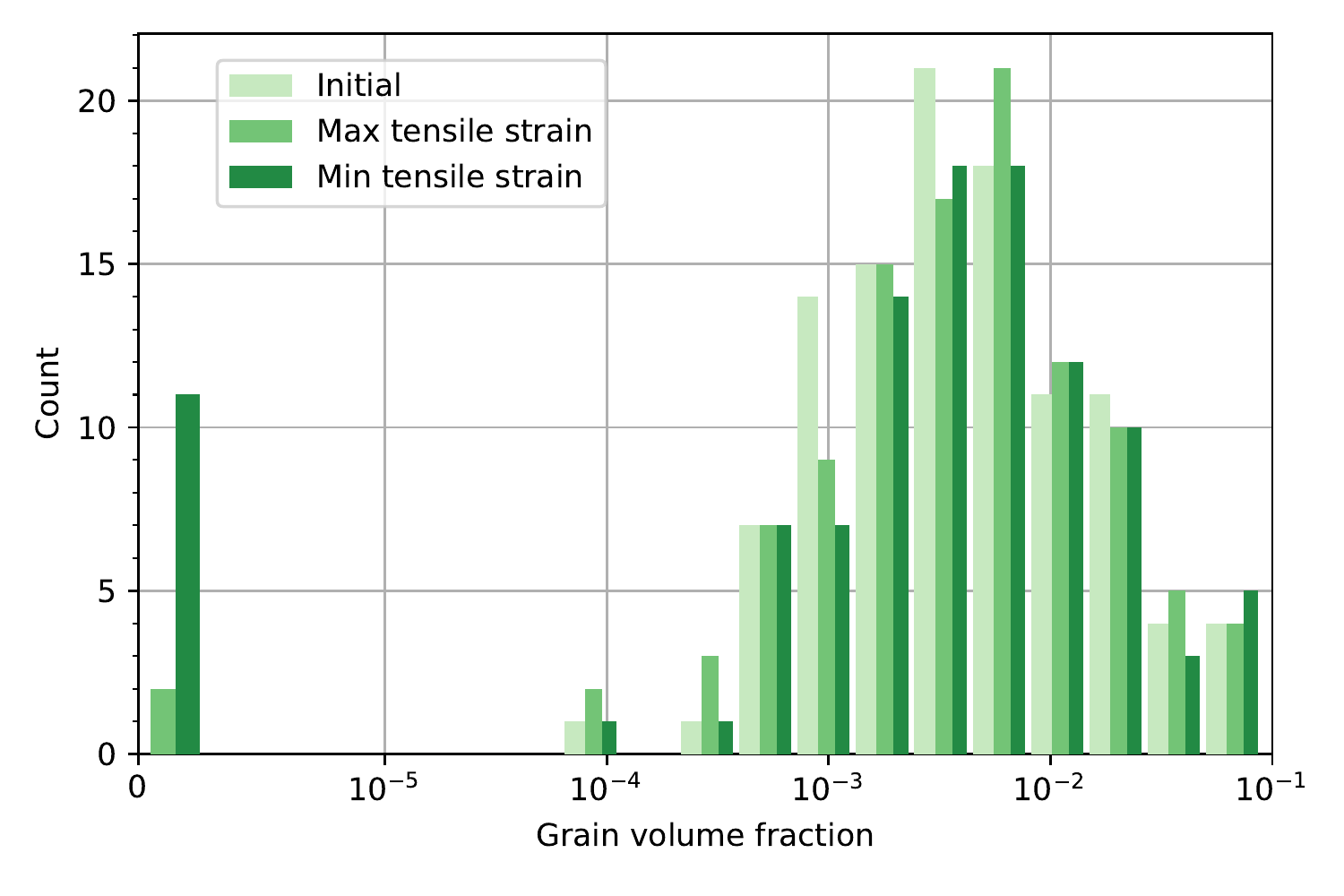}
    \caption{Non-GBE microstructure}
    \label{fig:hist_ngbe_lg}
  \end{subfigure}%
  \begin{subfigure}{0.5\linewidth}
    \includegraphics[width=\linewidth]{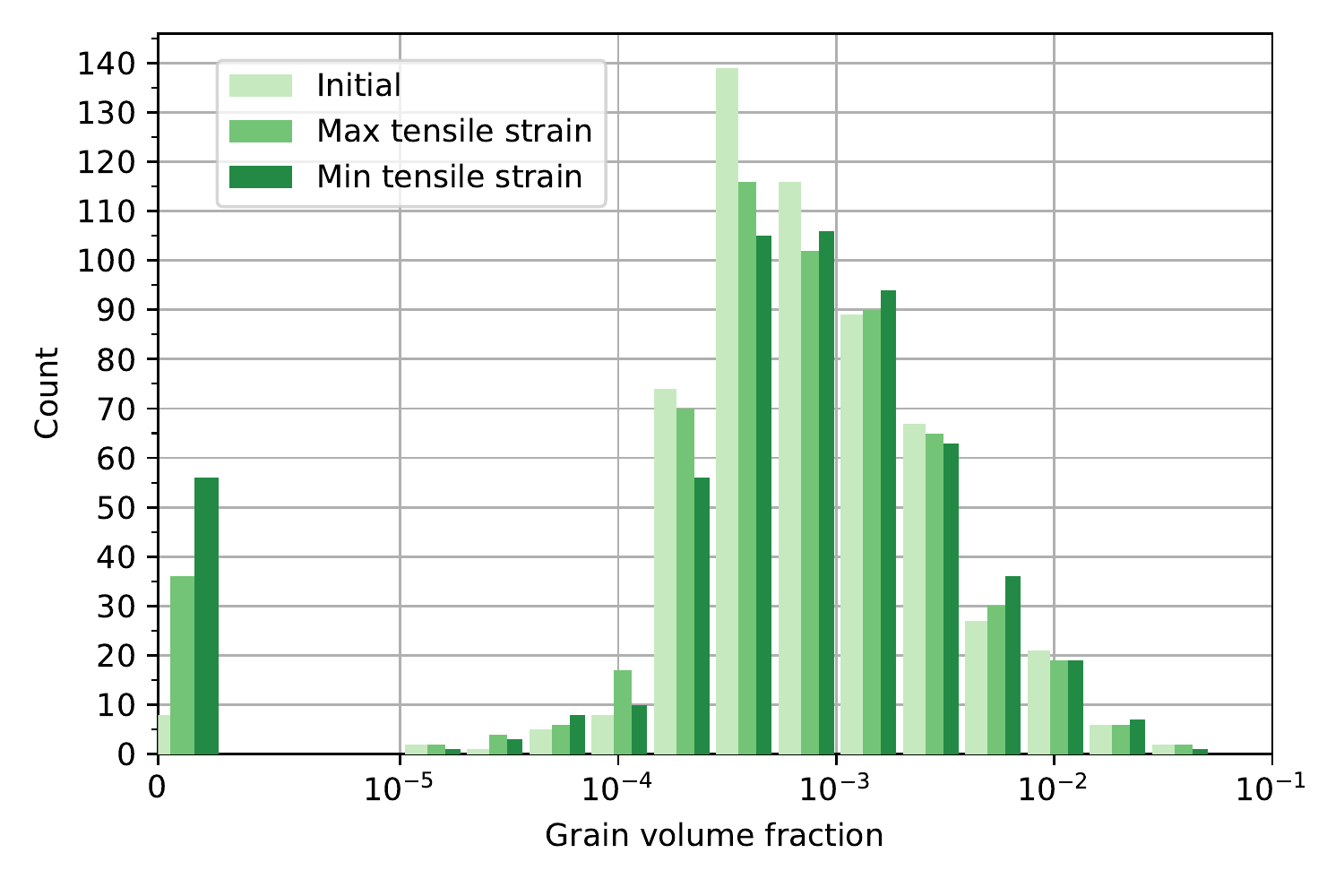}
    \caption{GBE microstructure}
    \label{fig:hist_gbe_lg}
  \end{subfigure}
  \caption{
      Evolution of grain size distribution for initial (light), maximum tension (medium) and maximum compression (dark).
      Results are plotted on a sym-log scale (linear from 0-$10^-5$) to include consumed (V=0) grains.
    }
\end{figure}

  As with the NRC ``random'' case in the previous section, all tests are run without recreation (grains with zero volume are locked and not allowed to grow back).
  The distribution of grain sizes (volume fractions) for the ``large'' cases of both GBE and non-GBE give insight into the evolution of the microstructure (\cref{fig:hist_ngbe_lg,fig:hist_gbe_lg}).
  In both cases, the number of zero-volume grains starts at zero and increases during the loading cycle.
  As a result, the average grain volume fraction increases slightly, indicated by a visible (though small) shift towards the right as the loading cycle increases.
  This behavior is qualitatively indicative of coarsening, and is a natural consequence of the no-grow-back condition.
  (Interestingly, coarsening to a lesser degree still occured even without this condition.)

\begin{figure}[h]
  \begin{subfigure}{0.33\linewidth}
    \includegraphics[width=\linewidth]{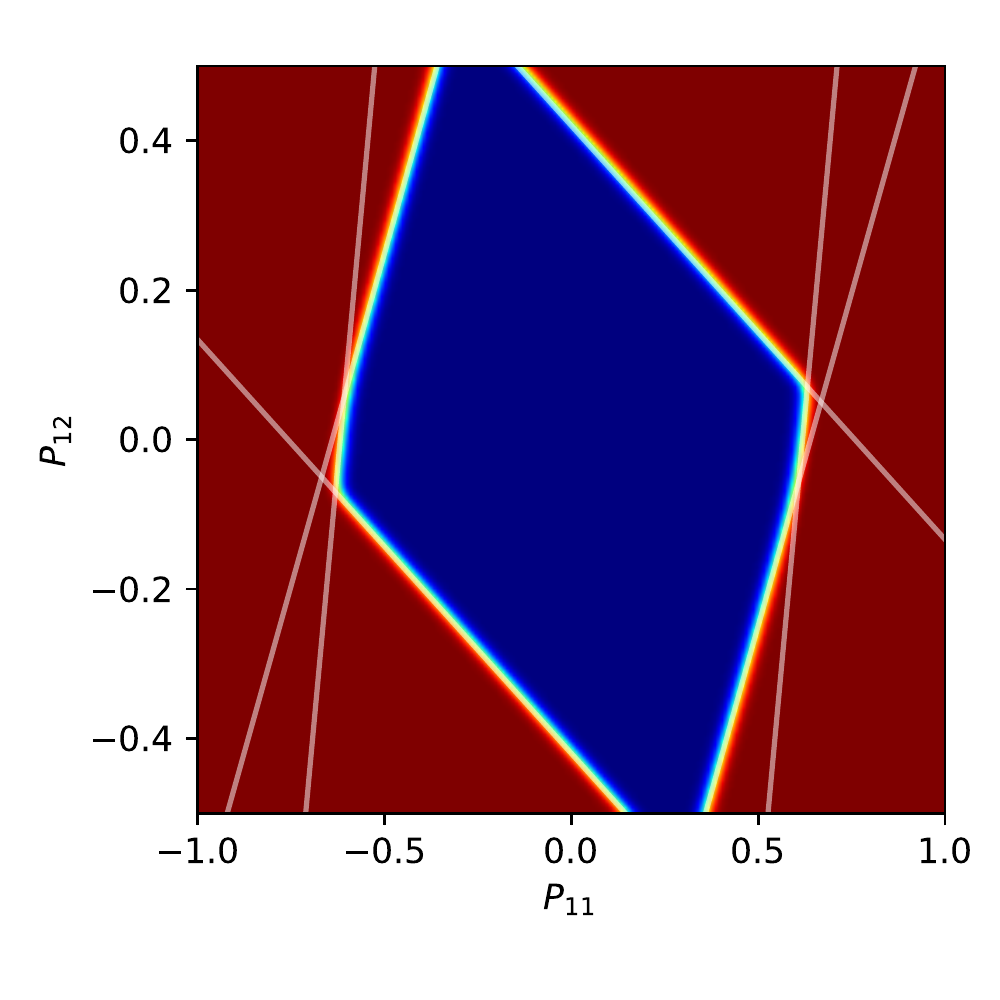}
    \caption{Non-GBE - Small}
    \label{fig:GrainData_GB_NONeng_Small_ys}
  \end{subfigure}%
  \begin{subfigure}{0.33\linewidth}
    \includegraphics[width=\linewidth]{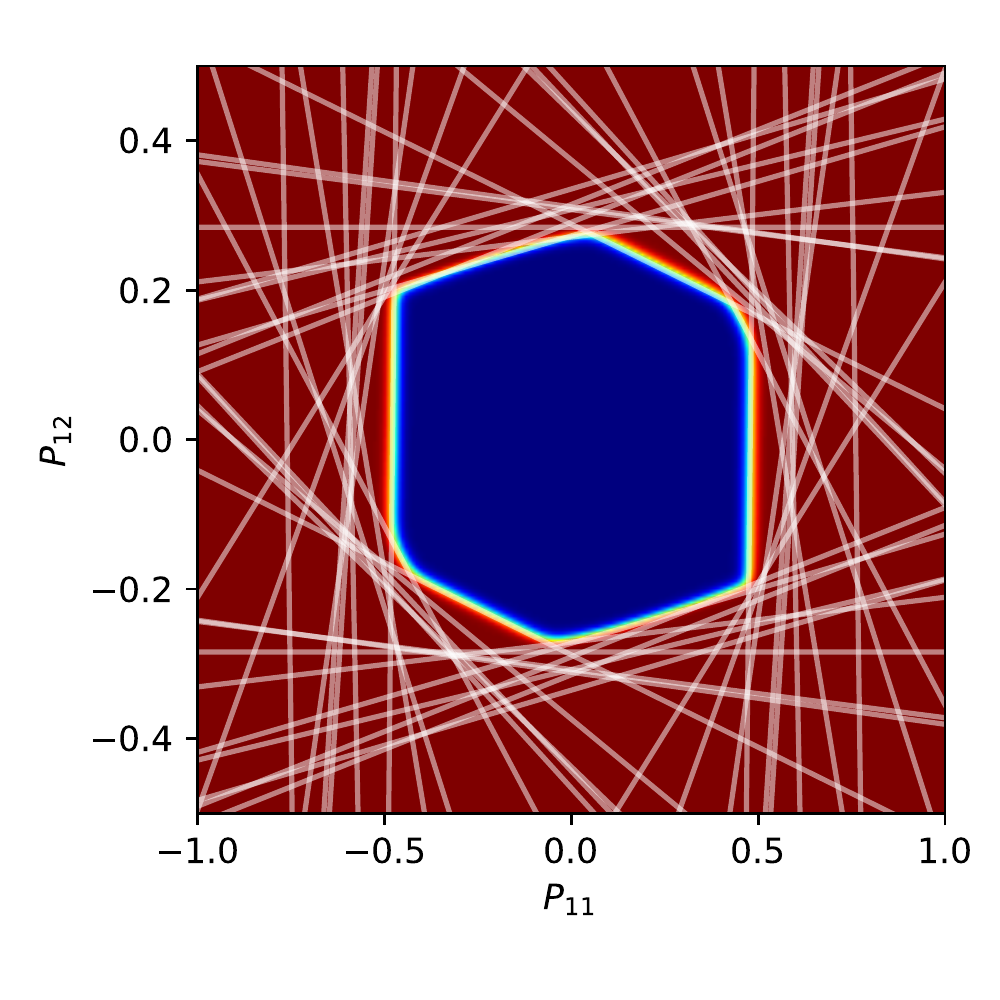}
    \caption{Non-GBE - Medium}
    \label{fig:GrainData_GB_NONeng_Medium_ys}
  \end{subfigure}%
  \begin{subfigure}{0.33\linewidth}
    \includegraphics[width=\linewidth]{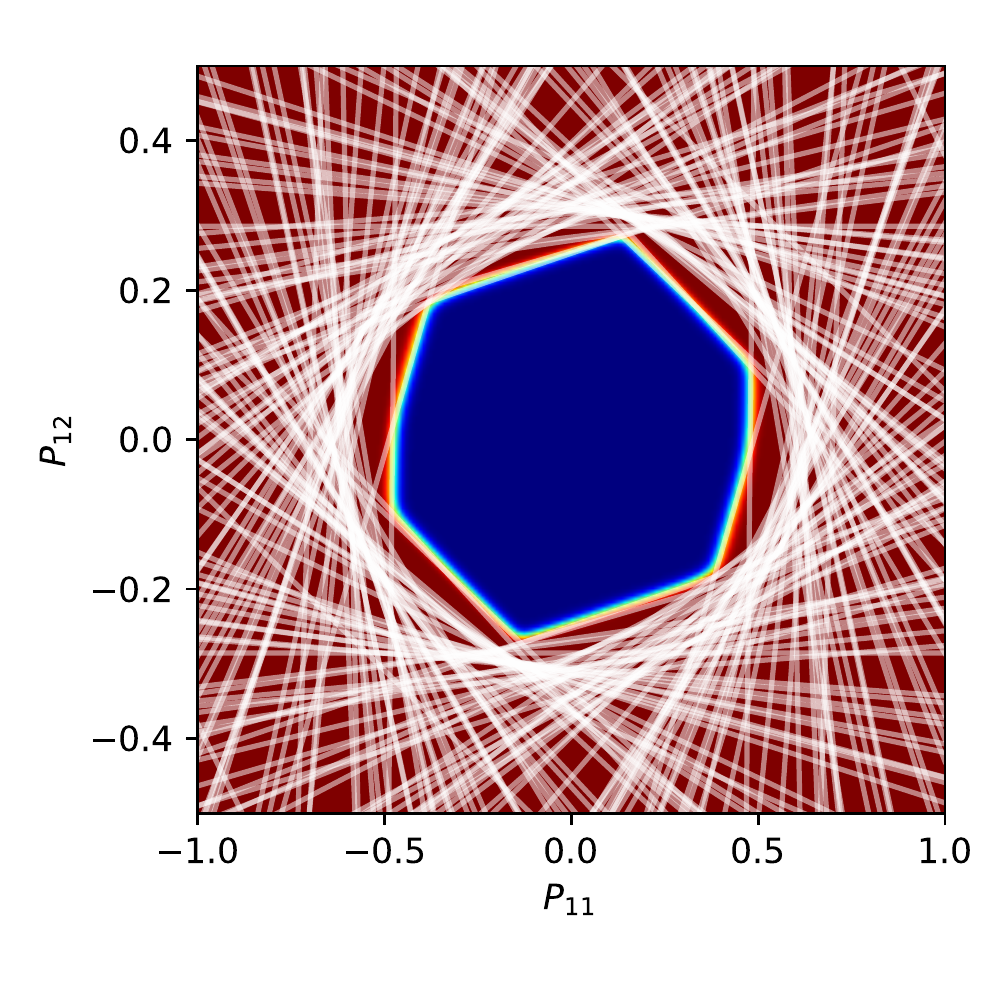}
    \caption{Non-GBE - Large}
    \label{fig:GrainData_GB_NONeng_Large_ys}
  \end{subfigure}

  \begin{subfigure}{0.33\linewidth}
    \includegraphics[width=\linewidth]{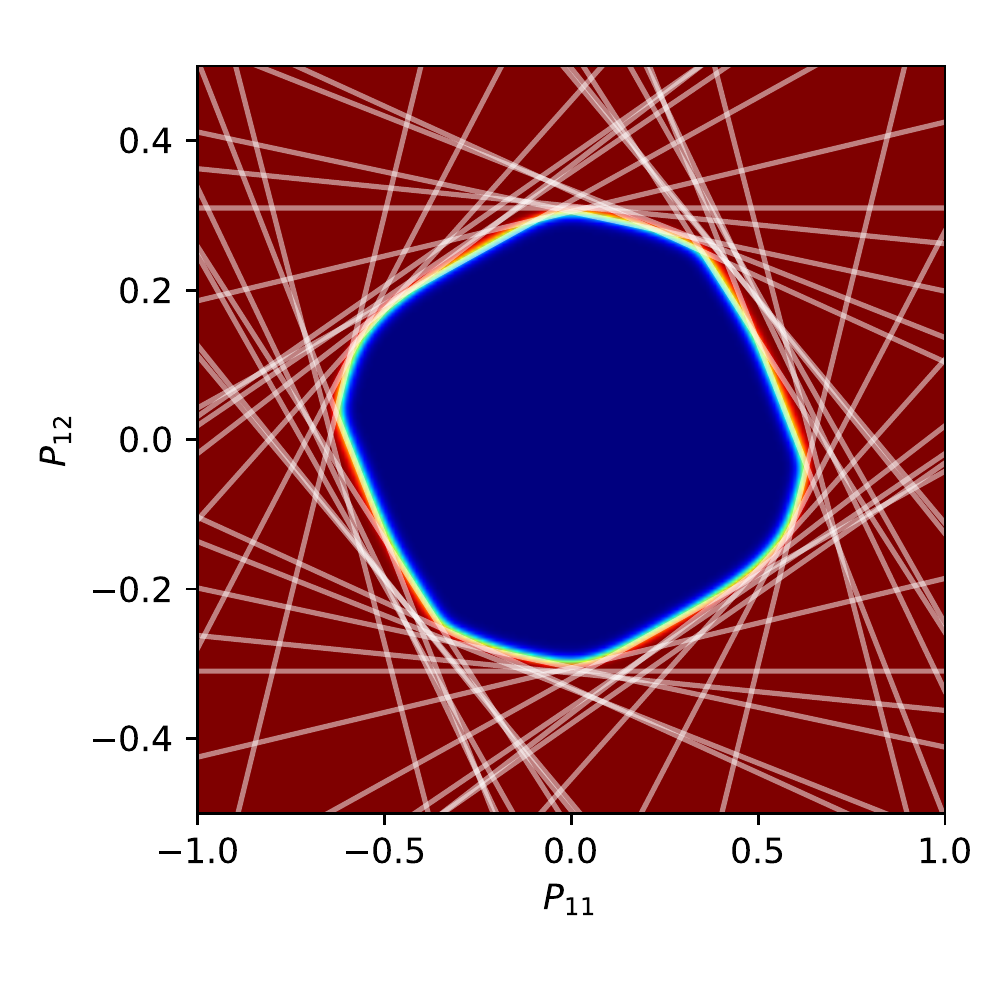}
    \caption{GBE - Small}
    \label{fig:GrainData_GBeng_Small_ys}
  \end{subfigure}%
  \begin{subfigure}{0.33\linewidth}
    \includegraphics[width=\linewidth]{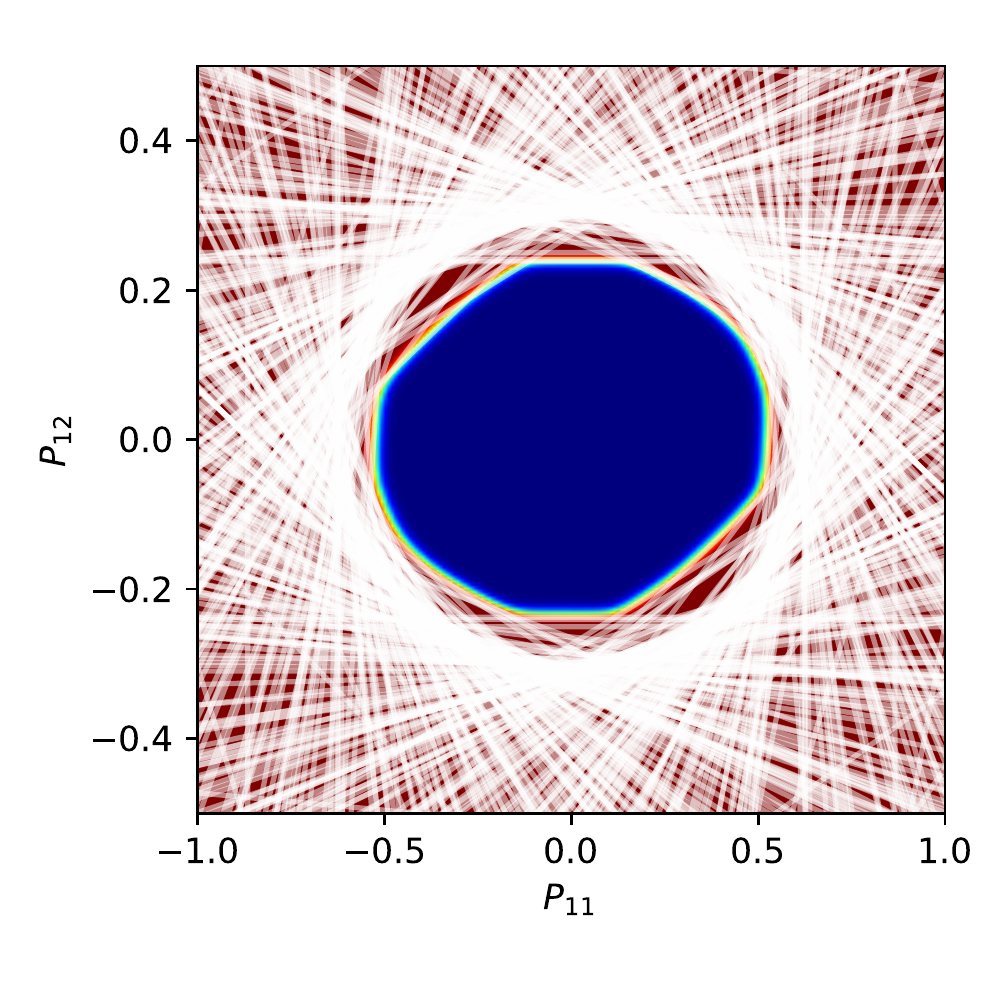}
    \caption{GBE - Medium}
    \label{fig:GrainData_GBeng_Medium_ys}
  \end{subfigure}%
  \begin{subfigure}{0.33\linewidth}
    \includegraphics[width=\linewidth]{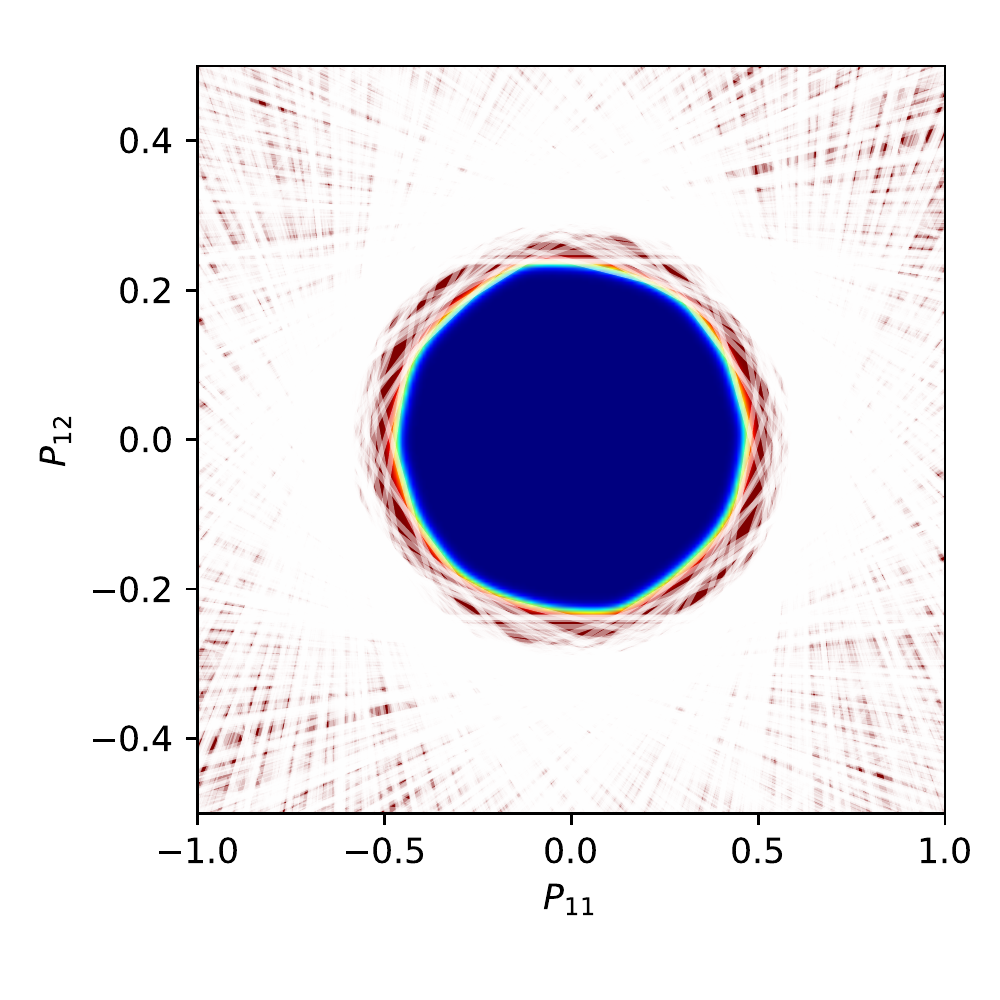}
    \caption{GBE - Large}
    \label{fig:GrainData_GBeng_Large_ys}
  \end{subfigure}
  \caption{Yield surfaces for GBE and Non-GBE microstructures}
\end{figure}

The similarities in the stress strain relations of the large samples are reflected in the yield surfaces, while the differences in the small samples are also noticeable.
The small sample size for both GBE and non-GBE produces a complicated yield surface (\cref{fig:GrainData_GB_NONeng_Small_ys,fig:GrainData_GBeng_Small_ys}), but approach nearly isotropic (circular/elliptical) shapes as the number of grains becomes large (\cref{fig:GrainData_GB_NONeng_Large_ys,fig:GrainData_GBeng_Large_ys}).  
With many grains, the yield surface resembles the J2 yield surface \cite{gu2009finite}, apparently converging once accounting for more than a couple hundred grains.
As with the previous cases, the white lines correspond to individual grain yield surfaces. 
As more grains are accounted for, many of the twin lines are offset by about 10\% of the yield surface radius, forming an apparent second yield surface.
Apparently, the boundaries with large shear coupling factors and low dissipation energies are responsible for the initial inner yield surface (\cref{fig:GrainData_GBeng_Large_ys}) and moderate shear coupling factors and dissipation energies create a second yield surface, which are the majority of GBs.
As explored with examination of the twin GBs and random GBs (\cref{fig:EBSD_twin,fig:EBSD_random}), large shear coupling factors with low dissipation energies are most likely to be twin boundaries; and therefore much of the plastic deformation it due to twin boundaries.

The yield surfaces from NP show similarities to other experimental and theoretical models of yield surfaces of non-work harden polycrystalline materials \cite{zattarin2004numerical}. 
The effect of texture evolution was found in \cite{zattarin2004numerical} to play an insignificant role for yield surface evolution.
Instead work hardening provides an appropriate tool to simulate the effect of large microstructures past yield.
However, \cite{watanabe1999control} found texture evolution to have a significant effect on a GBE Al-Li alloy causing weakening from an increase in the proportion of random boundaries at large deformations. 
With proper calibration, the hardening potentials play the same role as work hardening in crystal plasticity. 

To summarize, the implications from this example are threefold.
First, there is an obvious size effect in NP, where increased numbers of grains leads to increased ductility.
This is consistent with the inverse Hall-Petch effect, but indicates that crystal plasticity must be included to also capture the normal Hall-Petch effect.
Second, the plastic behavior exhibits a strong dependency on the microstructure, as evidenced by the variation in the yield surface and plastic hysteresis, for relatively small numbers of grains (1-150). 
Finally, for very large numbers of grains ($>$150), NP yield behavior approximates J2, and the plastic flow is governed by the hardening model.
Therefore, we conclude that NP is most effective at the 1-150 grain range, and can be suitably approximated by simpler plastic models at larger scales.

\section{Discussion}

In this work, a new approach is presented for modeling plasticity accommodated by microstructure evolution through the motion of grain boundaries.
Microstructure information is encoded through the use of a graph structure, where grains, quasigrains, and boundaries are stored on graph vertices and edges.
Mechanical equilibrium and the principle of minimum dissipation potential are used to derive governing equations for the vertex and edge quantities.
The result is a constitutive model that models the response of a microstructure network, in response to an applied local deformation or stress, through the flow of volumes along the graph.
Even though full topological transitions (representing creation or deletion of edges and/or vertices in the graph) are not fully explored here, simplistic coarsening can be modeled by locking zero-volume grains.
It was demonstrated that the model is able to capture the effects of microstructure, particularly when considering a relatively small number (1-150) of boundaries.
However, the model is simple enough to be implemented at the material point method in a larger continuum mechanics simulation.

As with all reduced-order models, NP makes a number of simplifications in order to achieve tractability.
The key assumptions are reviewed here.
(1) Weak compatibility: deformations are grain-wise constant, and must average to the overall deformation. 
As a result, stress is constant for each grain, and local high-resolution stress concentrations are not resolved.
(2) Planar boundaries: boundaries are parameterized by a single value.
This necessitates the use of hardening mechanisms to approximate the effects of non-planar motion (geometric hardening).
(3) Boundary-mediated plasticity only: in the present formulation, dislocation-mediated plasticity is not considered.
To incorporate crystal plasticity only requires that the grain-wise eigenstrains be evolved according to a crystal plasticity flow rule.
The graph structure of NP provides a suitable framework for combining the effects of crystal plasticity and grain boundary plasticity.
This is left to future work.

  It must also be noted that, except for the atomistic comparisons for STGBs in \cref{sec:ex_stgb}, the present work does not include validation.
  The results serve as a a proof-of-concept and demonsration of versatility rather than a comprehensive constitutive model.
  Therefore, we do not promote the present work as a validated material model.
  Rather, we present GB network plasticity as a framework allowing for the combination of many different plasticity models together, while simultaneously facilitating microstructure evolution in a reduced-order fashion.
  NP necessarily includes a very large number of parameters - on the order of dozens per grain and per boundary.
  All of the parameters have physical interpretability, but there is simply insufficient data (atomistic or experiment) for them to be reasonably calibrated, meaning that NP cannot yet be completely validated or used in a predictive fashion.
  However, the structure provided by NP may serve to motivate the generation of robust datasets that will lead to robust, multi-scale structre-property prediction.

\section{Acknowledgements}

DB acknowledges partial support from Lawrence Berkeley National Laboratory, grant \# 7645776.
BR acknowledges partial support from the National Science Foundation, grant \# MOMS-2341922.
This work used the INCLINE cluster at the University of Colorado Colorado Springs. 
INCLINE is supported by the National Science Foundation, grant \#2017917.
The authors wish to thank Professor Vinamra Agrawal (Auburn University) and Professor Nikhil Admal Chandra (University of Illinois Urbana-Champaign) for insightful discussions on this work.

\printbibliography

\appendix

\section{Mechanical equilibrium in network plasticity}\label{sec:mechanical_equilibrium}
The state of mechanical equilibrium in NP is determined by the calculation of grain-wise and quasigrain-wise deformation gradients that minimize the Helmholtz free energy while maintaining weak compatibility. 
That is, the Helmholtz free energy for a prescribed deformation gradient $\F$ is 
\begin{equation}
    \A^*(\F) = \sup_{\mathbf{\Lambda}} \inf_{\{\F_p\},\{\F_{pq}\}}\frac{1}{V}\Bigg[
    \sum_{p\in\verts}V_p\A(\F_p) + \sum_{pq\in\edges}|V_{pq}|\A(\F_{pq})
    - \mathbf{\Lambda}:
    \Bigg(
    \sum_{p\in\verts}V_p\F_p + \sum_{pq\in\edges}|V_{pq}|\F_{pq} - V\F)
    \Bigg)
    \Bigg]
\end{equation}
Stationarity for $\F_p$, $\F_{pq}$, and $\mathbf{\Lambda}$ yields
\begin{align}\label{eq:mech_eq_general}
    \P(\F^*_p) - \mathbf{\Lambda^*} &= \bm{0} &
    \P(\F^*_{pq}) - \mathbf{\Lambda^*} &= \bm{0} &
    \sum_{p\in\verts}V_p\F^*_p + \sum_{pq\in\edges}|V_{pq}|\F^*_{pq} - V\F^* &= \bm{0},
\end{align}
where the asterisk indicates optimality.
The system in \cref{eq:mech_eq_general} requires a constitutive relation for closure, and so we recall the special linearized elastic relations for shear coupled systems as described in \cref{Sec:MinDissPot}:
\begin{align}
    \mathbb{C}_{p}(\F_p - \I) - \mathbf{\Lambda}&= \bm{0} &
    \mathbb{C}_{pq}(\F_{pq} - \I - \sgn(h_{pq})\Fgb_{pq}) - \mathbf{\Lambda} &= \bm{0},
\end{align}
noting that the identity matrices $\I$ may be replaced with alternative eigenstrains as appropriate.
Insertion of these relations into the first stationarity conditions in \cref{eq:mech_eq_general} yields
\begin{align}
    \F_p  &= \I + \mathbb{C}_p^{-1}\mathbf{\Lambda} &
    \F_{pq} &= \I + \sgn(h_{pq})\Fgb_{pq} + \mathbb{C}_{pq}^{-1}\mathbf{\Lambda}.
\end{align}
Substitution of these into the stationarity condition for $\mathbf{\Lambda}$ (i.e. weak compatibility) gives
\begin{align}
    \sum_{p\in\verts}V_p\Big(\I + \mathbb{C}_p^{-1}\mathbf{\Lambda}\Big) 
    + \sum_{pq\in\edges}|V_{pq}|\Big(\I + \sgn(h_{pq})\Fgb_{pq} + \mathbb{C}_{pq}^{-1}\mathbf{\Lambda}\Big) = V\F,
\end{align}
which, once simplified and rearranged, becomes
\begin{align}
    \mathbf{\Lambda}
    =
    V\Bigg[\sum_{p\in\verts}V_p\mathbb{C}_p^{-1}
    + 
    \sum_{pq\in\edges}|V_{pq}|\mathbb{C}_{pq}^{-1}\Bigg]^{-1}\Bigg(\F-\I
    -
    \frac{1}{V}\sum_{pq\in\edges}V_{pq}\Fgb_{pq}\Bigg).
\end{align}
Recalling the definition of $\langle\mathbb{C}\rangle$ in \cref{eq:average_c}, as well as the original two stationarity conditions, shows that
\begin{align}
  \P^* = \P_p^* = \P_p(\F^*_p) = \P_{pq}^* =\P_{pq}(\F^*_{pq}) = \mathbf{\Lambda} = \langle\mathbb{C}\rangle \Bigg((\F-\I)
    -
    \frac{1}{V}\sum_{rs\in\edges}V_{rs}\Fgb_{rs}\Bigg)
\end{align}
Substituting, the optimal grain-wise and quasigrain-wise deformation gradients are given by
\begin{gather}
    \F^*_p
    =
    \I + 
    \mathbb{C}^{-1}_p
    \langle\mathbb{C}\rangle\Bigg(\F-\I
    -
    \frac{1}{V}\sum_{rs\in\edges}V_{rs}\Fgb_{rs}\Bigg) \label{eq:fp_optimal}\\
    \F^*_{pq}
    =
    \I + \Fgb_{pq} +  
    \mathbb{C}^{-1}_{pq}
    \langle\mathbb{C}\rangle\Bigg(\F-\I
    -
    \frac{1}{V}\sum_{rs\in\edges}V_{rs}\Fgb_{rs}\Bigg) \label{eq:fpq_optimal}
\end{gather}

\section{Derivation of driving force terms}\label{sec:derivation_driving_force_terms}
In this section, we are interested in the variation of $\F^*_{p\in\verts}$, $\F^*_{pq\in\edges}$ with respect to interface positions $h_{pq\in\edges}$.
Recall that the asterisk indicates optimality and, consequently, dependence upon the prescribed deformation gradient $\F$.
Using the product rule, the derivatives of the optimal deformation gradients under an applied load are:
\begin{gather}
    \frac{\partial\F^*_r}{\partial h_{pq}}
    =
    -\mathbb{C}^{-1}_r
    \langle\mathbb{C}\rangle
    \Bigg(
    \frac{a_{pq}}{V}
    \Fgb_{pq}
    \Bigg)
    +
    \frac{\partial}{\partial h_{pq}}(\mathbb{C}^{-1}_r \langle\mathbb{C})
    \rangle\Bigg(\F-\I
    -
    \frac{1}{V}\sum_{tu\in\edges}V_{tu}\Fgb_{tu}\Bigg)
    \\
    \frac{\partial\F^*_{rs}}{\partial h_{pq}}
    = -
    \mathbb{C}^{-1}_{rs}
    \langle\mathbb{C}\rangle\Bigg(
    \frac{a_{pq}}{V}
    \Fgb_{pq}\Bigg)
    +
    \frac{\partial}{\partial h_{pq}}(\mathbb{C}^{-1}_{rs} \langle\mathbb{C}\rangle) 
    \Bigg(\F-\I
    -
    \frac{1}{V}\sum_{tu\in\edges}V_{tu}\Fgb_{tu}\Bigg)
\end{gather}
Now the derivatives of the averaged deformation gradient must be computed.
It can be shown from the definition of $\langle\mathbb{C}\rangle$ that the derivative of the inverse reduces to
\begin{align}
  \frac{\partial}{\partial h_{pq}}\langle\mathbb{C}\rangle^{-1} = \frac{\partial}{\partial h_{pq}}\frac{\left(V_p\mathbb{C}^{-1}_p + V_q\mathbb{C}^{-1}_q + |V_{pq}|\mathbb{C}^{-1}_{pq}\right)}{V}.
\end{align}
Applying the volume flow rules results in the relationship
\begin{equation}
\label{average_cinv_dhpq_calc}
     \frac{\partial}{\partial h_{pq}} \langle\mathbb{C}\rangle^{-1} = \frac{a_{pq}}{V}(\mathbb{C}^{-1}_q - \mathbb{C}^{-1}_p )
\end{equation}
Applying the matrix identity $\frac{\partial A^{-1}}{\partial x} = -A^{-1}\frac{\partial A}{\partial x} A^{-1}$, gives
\begin{align}
  \frac{\partial}{\partial h_{pq}}\langle\mathbb{C}\rangle
  =
  -\langle\mathbb{C}\rangle\frac{\partial\langle\mathbb{C}\rangle^{-1}}{\partial h_{pq}}\langle\mathbb{C}\rangle
  =
  -\frac{a_{pq}}{V}\langle\mathbb{C}\rangle(\mathbb{C}_q^{-1}-\mathbb{C}_p^{-1})\langle\mathbb{C}\rangle
\end{align}
Substituting into the above yields
\begin{gather}
    \frac{\partial\F^*_r}{\partial h_{pq}}
    =
    -\mathbb{C}^{-1}_r
    \langle\mathbb{C}\rangle
    \Bigg(
    \frac{a_{pq}}{V}
    \frac{\partial|V_{pq}|}{\partial h_{pq}}\Fgb_{pq}
    \Bigg)
    -
    \frac{a_{pq}}{V}\mathbb{C}^{-1}_r\langle\mathbb{C}\rangle(\mathbb{C}_q^{-1}-\mathbb{C}^{-1}_p)\underbrace{\langle\mathbb{C}\rangle
    \Bigg(\F-\I
    -
    \frac{1}{V}\sum_{tu\in\edges}V_{tu}\Fgb_{tu}\Bigg)}_{\P^*}
    \\
    \frac{\partial\F^*_{rs}}{\partial h_{pq}}
    = -
    \mathbb{C}^{-1}_{rs}
    \langle\mathbb{C}\rangle\Bigg(
    \frac{a_{pq}}{V}
    \frac{\partial|V_{pq}|}{\partial h_{pq}}\Fgb_{pq}\Bigg)
    -
    \frac{a_{pq}}{V}\mathbb{C}^{-1}_r\langle\mathbb{C}\rangle(\mathbb{C}_q^{-1}-\mathbb{C}^{-1}_p)\underbrace{\langle\mathbb{C}\rangle
    \Bigg(\F-\I
    -
    \frac{1}{V}\sum_{tu\in\edges}V_{tu}\Fgb_{tu}\Bigg)}_{\P^*}
\end{gather}
which simplifies, finally, to 
\begin{gather}
    \frac{\partial\F^*_r}{\partial h_{pq}}
    =
    -
    \frac{a_{pq}}{V}
    \mathbb{C}^{-1}_r
    \langle\mathbb{C}\rangle
    \Big[
    \Fgb_{pq}
    +
    (\F^*_q-\F^*_p)\Big]
    \\
    \frac{\partial\F^*_{rs}}{\partial h_{pq}}
    =
    -
    \frac{a_{pq}}{V}
    \mathbb{C}^{-1}_{rs}
    \langle\mathbb{C}\rangle
    \Big[
    \Fgb_{pq}
    +
    (\F_q^*-\F^*_p)
    \Big]
\end{gather}

\end{document}